\documentclass[english]{article}
\usepackage[T1]{fontenc}
\usepackage{color}
\usepackage{graphicx}
\usepackage{amssymb}

\makeatletter

\usepackage{a4}
\input{epsfig.sty}
\def\fnum@table{\tablename~{\bf\thetable}}
\def\fnum@figure{\figurename~{\bf\thefigure}}
\def\tablename{\footnotesize{\bf Table}}
\def\figurename{\footnotesize{\bf Figure}}

\def\be{\begin{equation}}
\def\ee{\end{equation}}

\usepackage{babel}
\makeatother
\begin{document}

\title{\textbf{Enhanced Pomeron diagrams: re-summation of unitarity cuts}}

\author{\textbf{S. Ostapchenko} \\
\textit{\small Forschungszentrum Karlsruhe, Institut f\"ur} 
\textit{\small Kernphysik, 76021 Karlsruhe, Germany}\\
\textit{\small D.V. Skobeltsyn Institute of Nuclear Physics, Moscow
State University, 119992 Moscow, Russia} \textit{}\\
}

\maketitle
\begin{center}\textbf{\large Abstract}\end{center}{\large \par}

Unitarity cuts of enhanced Pomeron diagrams are analyzed in the framework
of the Reggeon Field Theory. Assuming the validity of the 
Abramovskii-Gribov-Kancheli cutting rules, we derive a complete set
 of cut non-loop enhanced graphs and observe important cancellations 
 between  certain sub-classes of the latter.
 We demonstrate  also how the present method can be generalized 
 to take into consideration Pomeron loop contributions.

\section{Introduction\label{intro.sec} }

Even nowadays, forty years after the Reggeon Field Theory (RFT) \cite{gri68}
has been proposed, it is widely applied for the description of high
energy hadronic and nuclear interactions. Partly, this is due to the fact
 that a number of important results of the
old RFT remain also valid in the perturbative BFKL Pomeron calculus
\cite{glr}. Thus, RFT remains a testing laboratory for novel
approaches, prior to their realization within more complicated BFKL
framework. On the other hand, a perturbative treatment of peripheral
hadronic collisions still remains a challenge, the processes being
dominated by {}``soft''  parton physics. Hence, when describing the
high energy behavior of total and diffractive hadronic cross sections,
calculating probabilities of large rapidity gap survival (RGS) in hadronic
final states, or developing general purpose  Monte Carlo (MC) generators,
one  applies the Pomeron phenomenology
 \cite{kai82,diflevin,difkmr,bond01,bor05,kmr07,wer93}.

\begin{figure}[htb]
\begin{center}\includegraphics[%
  width=7cm,
  height=3.5cm]{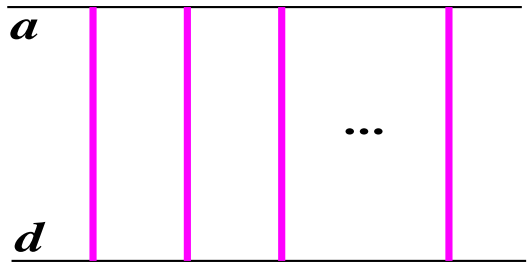}\end{center}
\vspace{-3mm}
\caption{General contribution to hadron-hadron scattering amplitude
from multiple Pomeron exchanges (vertical thick lines).\label{multiple}}
\end{figure}

Nevertheless, in MC applications one usually restricts himself with the
 comparatively simple multi-channel eikonal scheme, where  elastic
 scattering amplitude is described by diagrams
of Fig.~\ref{multiple},%
 corresponding to independent Pomeron exchanges between the two 
hadrons\footnote{Here we neglect  energy-momentum correlations
between multiple re-scatterings \cite{bra90}.},
and can be expressed via the Pomeron eikonal $\chi_{ad}^{\mathbb{P}}$
as \cite{kai82}
\begin{eqnarray}
f_{ad}(s,b)=i\sum _{j,l}C_{a(j)}\, C_{d(l)}
\left[1-e^{-\lambda_{a(j)}\, \lambda_{d(l)}\,\chi_{ad}^{\mathbb{P}}(Y,b)}\right]\!,
\label{f_ad}
\end{eqnarray}
with $Y=\ln s$, $s$ and $b$ being c.m.~energy squared and impact
parameter for the scattering.
  A small imaginary part of $\chi_{ad}^{\mathbb{P}}$ 
can be neglected in the high energy limit.
Here $C_{a(j)}$ and $\lambda_{a(j)}$ are correspondingly
relative weights and relative strengths\footnote{Here one makes a simplifying
assumption that elastic scattering amplitudes for hadronic states $|a\rangle$
and  $|d\rangle$ and for their low mass inelastic excitations $|a^*\rangle$
and  $|d^*\rangle$ differ  only by the corresponding couplings to the Pomeron.
In general, one may also consider different profile shapes for such amplitudes
(see, e.g. \cite{kmr07}).}
 of diffraction eigenstates $|a_j\rangle$
of hadron $a$ in the multi-component scattering scheme \cite{kai82}:
\begin{eqnarray}
|a\rangle =\sum_j \sqrt{C_{a(j)}}\,|a_j\rangle \,,\label{a-j} 
\end{eqnarray}
with $\sum_{j}C_{a(j)}=1$, $\sum_{j}C_{a(j)}\lambda_{a(j)}=1$. 

The optical theorem allows one to obtain immediately  total hadron-hadron cross section:
\begin{eqnarray}
\sigma^{\rm tot}_{ad}(s)=2\int \!d^2b\;{\rm Im}f_{ad}(s,b)=2\sum _{j,l}C_{a(j)}\, C_{d(l)}
\int \!d^2b\;\left[1-e^{-\lambda_{a(j)}\, \lambda_{d(l)}\,\chi_{ad}^{\mathbb{P}}(Y,b)}\right]\!.
\label{sig-tot}
\end{eqnarray}
On the other hand, in order to derive partial cross sections for various hadronic final states,
one applies the  Abramovskii-Gribov-Kancheli (AGK) cutting procedure \cite{agk} to obtain
asymptotically non-negligible unitarity cuts of the elastic scattering
diagrams of Fig.~\ref{multiple}. Combining together contributions of cuts of certain
topologies, one can identify them with partial contributions of particular final states.

\begin{figure}[htb]
\begin{center}
\includegraphics[%
  width=15cm,
  height=4cm]{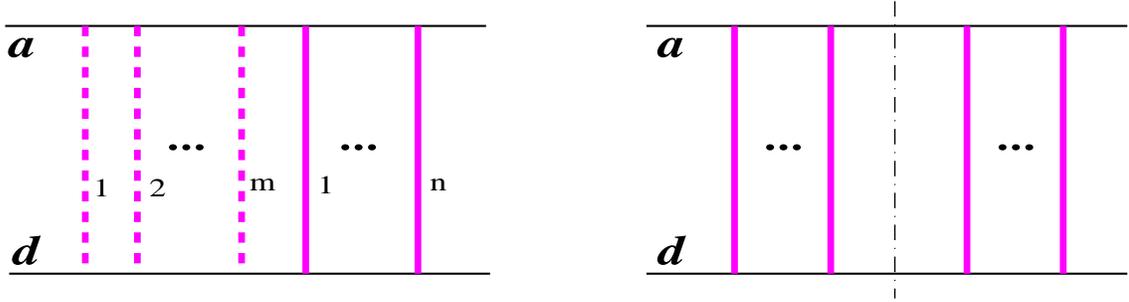}
\end{center}
\vspace{-3mm}
\caption{General contribution to multiple production cross section (left) and 
to the elastic and low mass diffraction cross sections (right). Cut  and uncut Pomerons
are shown by respectively dashed and solid vertical thick lines; the cut plane
is indicated when relevant by the dot-dashed line.\label{multcut}}
\end{figure}

For example, the so-called topological cross sections, corresponding to the interaction being
composed of $m\geq 1$  ``elementary'' particle production processes, 
are given by the contributions
of of graphs in Fig.~\ref{multcut}~(left),%
 with precisely $m$ Pomerons being cut, and with any number
$n$ of uncut ones \cite{kai82}: 
\begin{eqnarray}
\sigma^{(m)}_{ad}(s)=\sum_{n=0}^{\infty}\sum _{j,l}C_{a(j)}\, C_{d(l)}\int \!d^2b\;
\frac{\left[2\lambda_{a(j)}\, \lambda_{d(l)}\,\chi_{ad}^{\mathbb{P}}(Y,b)\right]^m}{m!}
\frac{\left[-2\lambda_{a(j)}\, \lambda_{d(l)}\,\chi_{ad}^{\mathbb{P}}(Y,b)\right]^n}{n!}
&&\nonumber \\
=\sum _{j,l}C_{a(j)}\, C_{d(l)}\int \!d^2b\;
\frac{\left[2\lambda_{a(j)}\, \lambda_{d(l)}\,\chi_{ad}^{\mathbb{P}}(Y,b)\right]^m}{m!}\;
 e^{-2\lambda_{a(j)}\, \lambda_{d(l)}\,\chi_{ad}^{\mathbb{P}}(Y,b)}\,.&&
\label{sig-(m)}
\end{eqnarray}

On the other hand, requiring the cut plane to pass between uncut Pomerons,
with at least one on either side of the cut, as shown in 
Fig.~\ref{multcut}~(right), one obtains
\begin{eqnarray}
\sigma^{(0)}_{ad}(s)=\sum _{j,l}C_{a(j)}\, C_{d(l)}\int \!d^2b\;
\left[1-e^{-\lambda_{a(j)}\, \lambda_{d(l)}\,\chi_{ad}^{\mathbb{P}}(Y,b)}
\right]^2\!,
\label{sig-(0)}
\end{eqnarray}
which can be further split into elastic and diffraction dissociation cross sections \cite{kai82}.
One can easily verify that the sum of (\ref{sig-(m)}) and (\ref{sig-(0)}) satisfies the 
$s$-channel unitarity relation:
\begin{eqnarray}
\sigma^{(0)}_{ad}(s)+\sum_{m=1}^{\infty}\sigma^{(m)}_{ad}(s)=\sigma^{\rm tot}_{ad}(s)\,.
\end{eqnarray}

\begin{figure}[htb]
\begin{center}
\includegraphics[%
  width=15cm,
  height=3.5cm]{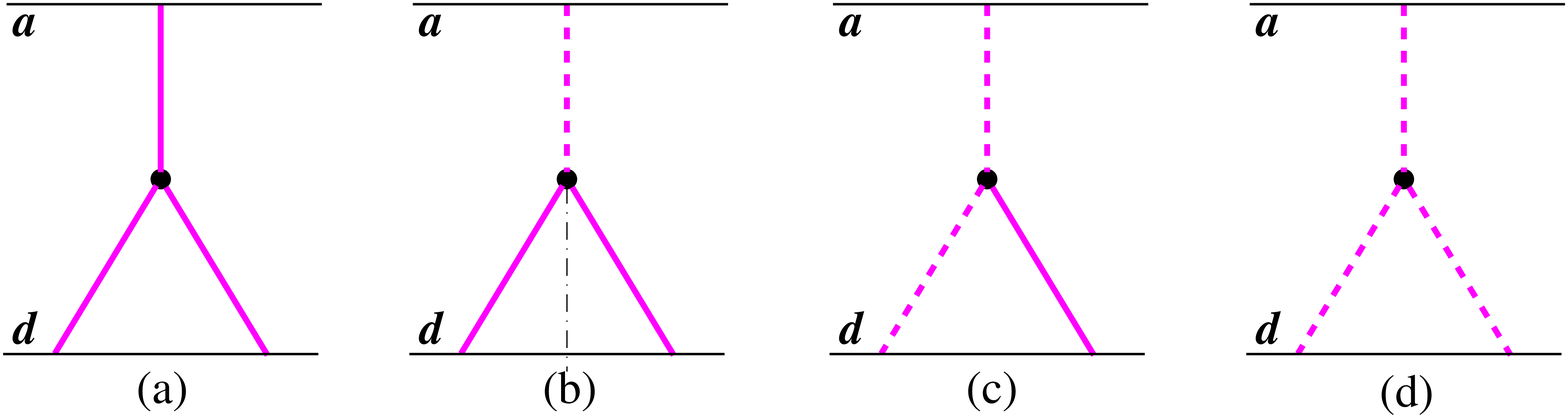}\end{center}
\vspace{-3mm}
\caption{Triple-Pomeron contribution to the elastic scattering amplitude (a)
and its AGK cuts:  high mass diffraction
 contribution  (b), screening correction to one cut Pomeron process (c), 
 and  ``cut Pomeron fusion'' process (d).\label{ycut}}
\end{figure}

However, the above-described scheme can  account only
 for low mass inelastic excitations of 
the projectile and target hadrons,
 where the integration over the masses of those
inelastic intermediate states can be performed irrespective the total
 c.~m.~energy $s$ for
the scattering \cite{kai82}.
 To treat high mass diffraction, one has to generalize the scheme, including
 the contributions of enhanced Pomeron diagrams, i.e.~to take
  Pomeron-Pomeron  interactions into account
   \cite{kan73,car74,bond01,bor05,kmr07}. 
  Moreover, such enhanced diagrams provide important absorptive
 corrections to the cross sections  (\ref{sig-tot}--\ref{sig-(0)}) and generate new final
 states of complicated topologies \cite{car74,bor05,kmr07,ost06,ost06a}.
  For example, cutting the simplest
 triple-Pomeron diagram of Fig.~\ref{ycut}~(a),%
 one obtains the projectile high mass diffraction
 contribution   of Fig.~\ref{ycut}~(b), a screening correction 
 to one cut Pomeron process
 (Fig.~\ref{ycut}~(c)), and a new ``cut Pomeron fusion'' process 
 of Fig.~\ref{ycut}~(d). With increasing
 energy, more complicated enhanced diagrams, with numerous multi-Pomeron
 vertices, become important. Thus, to obtain meaningful expressions for the contributions
 of various hadronic final states, one has to perform a re-summation of the whole series
 of the corresponding cut diagrams.
 
 A general procedure for the re-summation of enhanced diagram contributions to the elastic
 scattering amplitude has been proposed in \cite{ost06}. The goal of the present work is
 to apply the method for the re-summation of cut diagram contributions and to obtain a full
 set of the AGK-based unitarity cuts for the considered class
  of uncut enhanced Pomeron graphs.  Mainly, we  deal below
  with diagrams of ``net'' type, i.e. with arbitrary enhanced diagrams 
  which do not contain
  Pomeron ``loops'' (multi-Pomeron vertices connected 
  to each other by two or more Pomerons),
  although we shall demonstrate how the method can be generalized to include rather general
  Pomeron loop contributions. An analysis of the structure of final states corresponding
  to various unitarity cuts and an implementation of the approach in a hadronic MC generator
  is discussed elsewhere \cite{ost08}.
  
  The paper is organized as follows. In Section 2 we remind the basic results of the
  earlier works  \cite{ost06,ost06a} on the re-summation of uncut enhanced diagrams. In Section 3
  we analyze unitarity cuts of certain sub-graphs of general net diagrams
   (so-called
  ``net fans'') and  perform a re-summation of the cuts characterized by certain topologies
  of cut Pomerons. Next, in Section 4 we use those re-summed contributions as building
  blocks in the construction of the full set of cut non-loop enhanced diagrams, corresponding
  to the full discontinuity of the elastic scattering contributions of Section 2. Finally,
  in Section 5 we outline a generalization of  the present scheme 
  to include  Pomeron loop
  contributions. We conclude in Section 6.
  
\section{Uncut enhanced diagrams}
\begin{figure}[htb]
\begin{center}\includegraphics[%
  width=12cm,
  height=4cm]{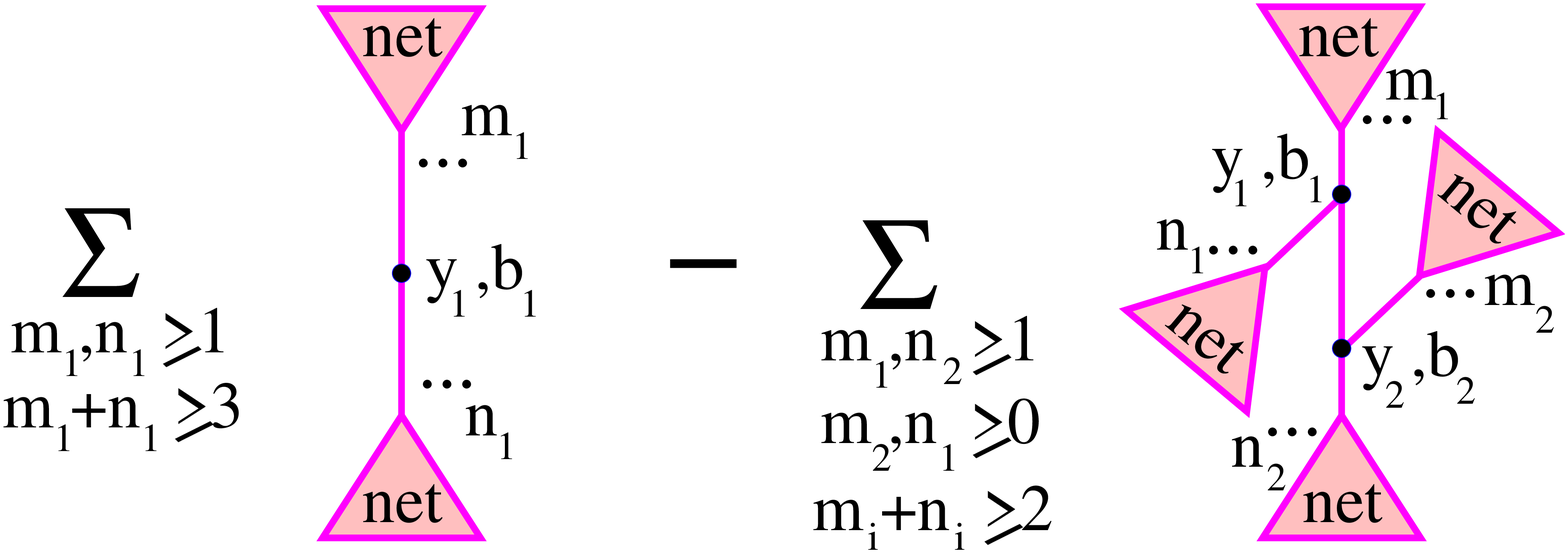}\end{center}
\vspace{-3mm}
\caption{Irreducible contributions of non-loop enhanced diagrams to elastic scattering
amplitude; the vertices $(y_i,b_i)$ are coupled to $m_i$ 
projectile and $n_i$ target ``net fans'', $i=1,2$;
$y_i$ and $\vec b_i$ define respectively rapidity and impact parameter positions
of multi-Pomeron vertices.\label{enh-full}}
\end{figure}

\begin{figure}[htb]
\begin{center}\includegraphics[%
  width=12cm,
  height=4cm]{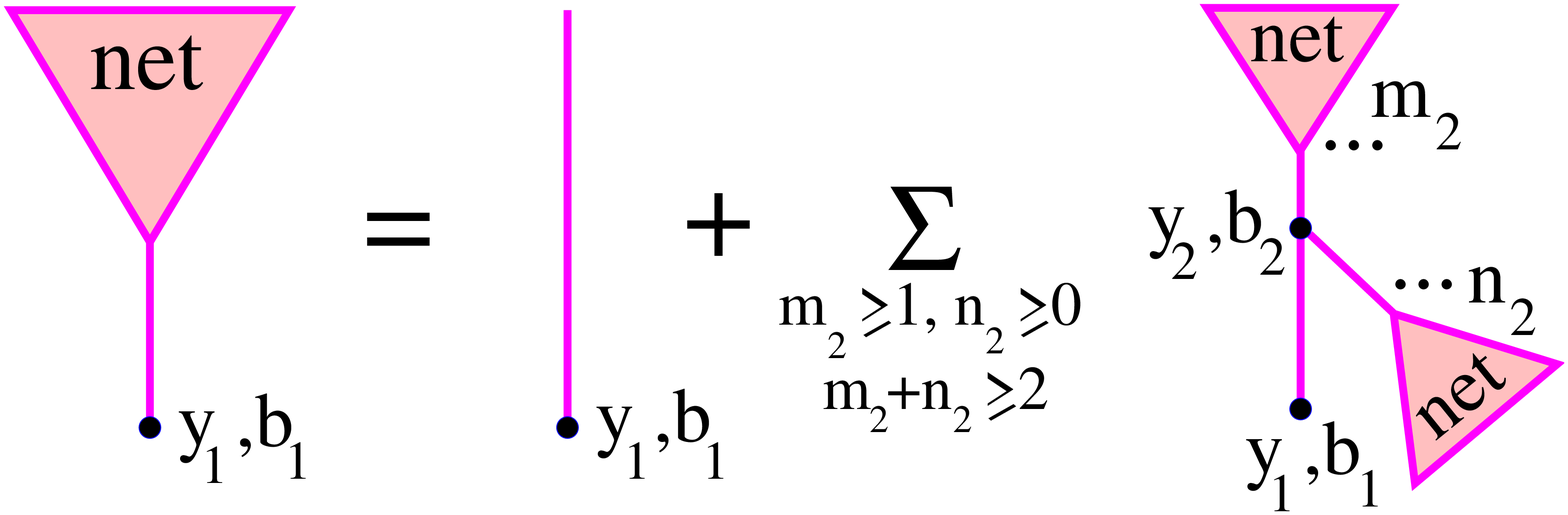}\end{center}
\vspace{-3mm}
\caption{Recursive equation for the {}``net fan'' contribution
 $\chi_{a(j)|d(l)}^{{\rm net}}(y_{1},\vec{b}_{1}|Y,\vec{b})$;
 $y_{1}$ and  $b_{1}$ are   rapidity and impact
 parameter distances between hadron $a$ and the vertex in the
 ``fan handle''.
\label{freve}}
\end{figure}

Taking Pomeron-Pomeron interactions into account, one has to consider multiple exchanges
of  coupled enhanced graphs, in addition to simple Pomeron
exchanges of Fig.~\ref{multiple}. The elastic scattering amplitude
can still be written in the usual multi-channel eikonal form (c.f.~(\ref{f_ad})):
\begin{eqnarray}
f_{ad}(s,b)=i\sum _{j,l}C_{a(j)}\, C_{d(l)}
\left[1-e^{-\lambda_{a(j)}\, \lambda_{d(l)}\,\chi_{ad}^{\mathbb{P}}(Y,b)
+\chi_{ad(jl)}^{{\rm enh}}(Y,b)}\right]\!,
\label{f_ad_enh}
\end{eqnarray}
where $\chi_{ad(jl)}^{{\rm enh}}$ stays for the eikonal contribution of irreducible enhanced 
graphs exchanged between  the diffraction eigenstates $|a_j\rangle$ and  $|d_l\rangle$ 
of hadrons $a$ and $d$. Restricting oneself with non-loop
enhanced diagrams, one can express  $\chi_{ad(jl)}^{{\rm enh}}$  
via the contributions
$\chi_{a(j)|d(l)}^{{\rm net}}$  of sub-graphs of certain structure,
 so-called ``net fans'', as shown in Fig.~\ref{enh-full} \cite{ost06,ost06a}.%
Here the ``net fan'' contributions $\chi_{a(j)|d(l)}^{{\rm net}}$
 are defined by the recursive equation of Fig.~\ref{freve}.%

In particular, assuming  eikonal structure of the vertices for 
the transition of $m$ into $n$ Pomerons \cite{car74}:
\begin{eqnarray}
g_{mn}=G\,\gamma_{\mathbb{P}}^{m+n}\,,\label{gmn}
\end{eqnarray}
with $r_{3{\mathbb{P}}}=G/(8\pi \gamma_{\mathbb{P}}^3)$ being
 the triple-Pomeron constant, the representations of
  Figs.~\ref{enh-full},~\ref{freve} yield~\cite{ost06,ost06a}: 
\begin{eqnarray}
\chi_{ad(jl)}^{{\rm enh}}(Y,b)=
G\int_{0}^{Y}\! dy_{1}\int\! d^{2}b_{1}\left\{ (1-e^{-\chi_{a|d}^{{\rm net}}(1)})
\,(1-e^{-\chi_{d|a}^{{\rm net}}(1)})
-\chi_{a|d}^{{\rm net}}(1)\;\chi_{d|a}^{{\rm net}}(1)\right.\nonumber \\
-G\int_{0}^{y_{1}}\! dy_{2}\int\! d^{2}b_{2}\;
\chi_{\mathbb{PP}}^{\mathbb{P}}(y_{1}-y_{2},|\vec{b}_{1}-\vec{b}_{2}|)\;
 \left[(1-e^{-\chi_{a|d}^{{\rm net}}(1)})
\,e^{-\chi_{d|a}^{{\rm net}}(1)}-\chi_{a|d}^{{\rm net}}(1)\right]
\nonumber \\
\left.\times\left[(1-e^{-\chi_{d|a}^{{\rm net}}(2)})\,
e^{-\chi_{a|d}^{{\rm net}}(2)}-\chi_{d|a}^{{\rm net}}(2)\right]\right\} \label{chi-enh}\\
\chi_{a(j)|d(l)}^{{\rm net}}(y_{1},\vec{b}_{1}|Y,\vec{b})
=\lambda_{a(j)}\,\chi_{a\mathbb{P}}^{\mathbb{P}}(y_{1},b_{1})
+G\int_{0}^{y_{1}}\! dy_{2}\int\! d^{2}b_{2}\;
\chi_{\mathbb{PP}}^{\mathbb{P}}(y_{1}-y_{2},|\vec{b}_{1}-\vec{b}_{2}|)\nonumber \\
\times\left\{ (1-e^{-\chi_{a(j)|d(l)}^{{\rm net}}\!(y_{2},\vec{b}_{2}|Y,\vec{b})})
\; e^{-\chi_{d(l)|a(j)}^{{\rm net}}\!(Y-y_{2},\vec{b}-\vec{b}_{2}|Y,\vec{b})}
-\chi_{a(j)|d(l)}^{{\rm net}}(y_{2},\vec{b}_{2}|Y,\vec{b})\right\} \!,\label{net-fan}
\end{eqnarray}
where $\chi_{a\mathbb{P}}^{\mathbb{P}}(y_{1},b_{1})$ is
 the eikonal for a Pomeron 
exchange between hadron $a$ and  the vertex  $(y_{1},\vec{b}_{1})$,
$y_{1}$ and  $b_{1}$ being  rapidity and impact
 parameter distances between hadron $a$ and that vertex, whereas the eikonal
$\chi_{\mathbb{PP}}^{\mathbb{P}}(y_{1}-y_{2},|\vec{b}_{1}-\vec{b}_{2}|)$
corresponds to a  Pomeron exchange between the vertices $(y_{1},\vec{b}_{1})$
and $(y_{2},\vec{b}_{2})$. In (\ref{chi-enh}) we used the abbreviations
 $\chi_{a|d}^{{\rm net}}(i)
 \equiv\chi_{a(j)|d(l)}^{{\rm net}}(Y-y_{i},\vec{b}-\vec{b}_{i}|Y,\vec{b})$,
$\chi_{d|a}^{{\rm net}}(i)\equiv
\chi_{d(l)|a(j)}^{{\rm net}}(y_{i},\vec{b}_{i}|Y,\vec{b})$, $i=1,2$.

\begin{figure}[htb]
\begin{center}\includegraphics[%
  width=15cm,
  height=3.cm]{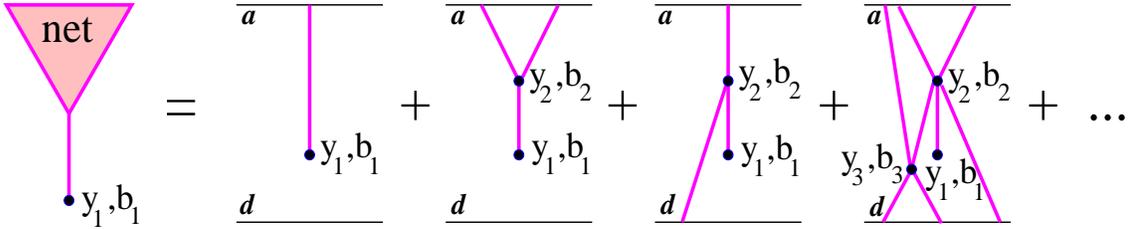}\end{center}
\vspace{-3mm}
\caption{Some examples of ``net fan'' diagrams.\label{net-examples}}
\end{figure}

The nickname ``net fan'' for the contribution 
$\chi_{a(j)|d(l)}^{{\rm net}}(y_{1},\vec{b}_{1}|Y,\vec{b})$
is because the Schwinger-Dyson equation of Fig.~\ref{freve} generates
Pomeron nets exchanged between hadrons $a$ and $d$, starting from a 
given vertex $(y_{1},\vec{b}_{1})$, some examples shown in Fig.~\ref{net-examples},%
and because this equation is formally similar to the usual fan
 diagram equation.
The latter can be recovered setting $n_2\equiv 0$ in Fig.~\ref{freve}.
In that case (\ref{net-fan})  reduces to
\begin{eqnarray}
\chi_{a(j)}^{{\rm fan}}(y_{1},{b}_{1})
=\lambda_{a(j)}\,\chi_{a\mathbb{P}}^{\mathbb{P}}(y_{1},b_{1})
+G\int_{0}^{y_{1}}\! dy_{2}\int\! d^{2}b_{2}\;
\chi_{\mathbb{PP}}^{\mathbb{P}}(y_{1}-y_{2},|\vec{b}_{1}-\vec{b}_{2}|)\nonumber \\
\times\left[ 1-e^{-\chi_{a(j)}^{{\rm fan}}\!(y_{2},{b}_{2})}
-\chi_{a(j)}^{{\rm fan}}(y_{2},{b}_{2})\right] \!.\label{fan}
\end{eqnarray}

In contrast to the fan contribution $\chi_{a(j)}^{{\rm fan}}$,
 which can be 
associated with parton density of a free hadron $a$ \cite{glr}, 
the ``net fan''
equation of  Fig.~\ref{freve}  accounts for absorptive corrections due to 
the re-scattering  on the partner hadron $d$ and corresponds to parton momentum
and impact parameter distribution which is probed during the interaction 
\cite{ost06,kmr07}. In the following, the Pomeron connected to the
 initial vertex  $(y_{1},\vec{b}_{1})$ in Fig.~\ref{freve} will be referred to 
 as the ``fan handle''.

In the representation of  Fig.~\ref{enh-full} for  enhanced diagram
contribution to the elastic scattering amplitude, the first graph in the 
r.h.s.~corresponds
to any number $m_1\geq 1$ of projectile ``net fans''
 $\chi_{a(j)|d(l)}^{{\rm net}}$ 
and any number $n_1\geq 1$ of target ones $\chi_{d(l)|a(j)}^{{\rm net}}$,
$m_1+n_1\geq 3$, which are coupled together in some ``central'' vertex, 
whereas the second
graph in the Figure is the double counting correction. Any diagram with
$n$ multi-Pomeron vertices is generated $n$ times by the first graph in the 
r.h.s.~of  Fig.~\ref{enh-full} (as there are $n$ choices for the
 ``central'' vertex),
from which $(n-1)$ contributions are subtracted by the second graph.

\section{Unitarity cuts of {}``net fans''\label{fans.sec} }

\begin{figure}[htb]
\begin{center}\includegraphics[%
  width=12.5cm,
  height=4cm]{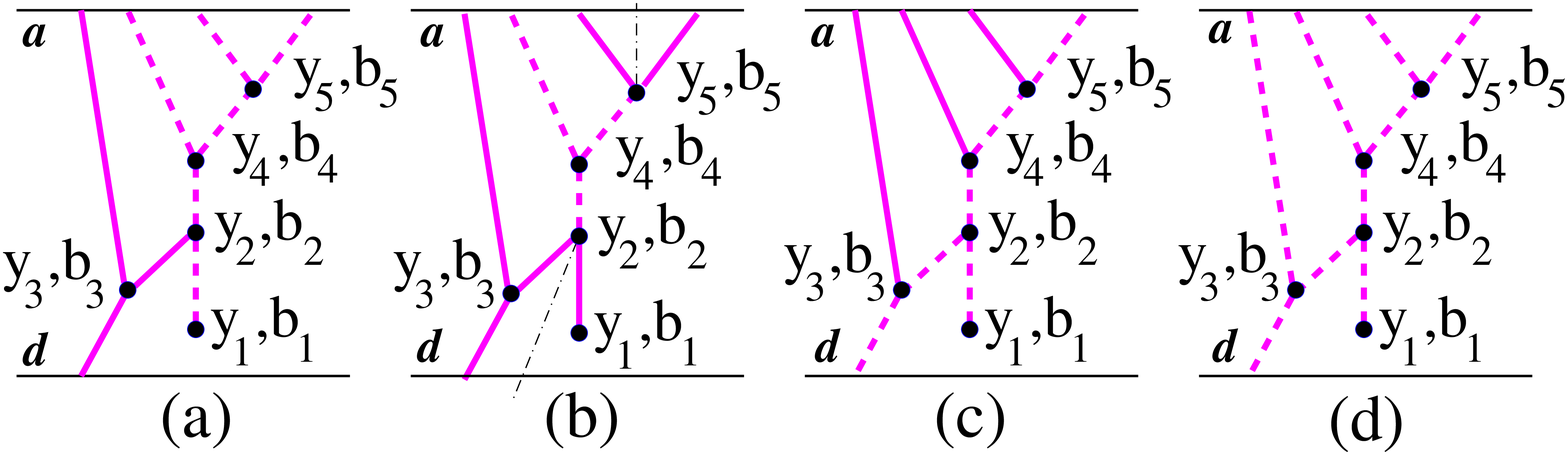}\end{center}
\vspace{-3mm}
\caption{Examples of graphs obtained by cutting the same projectile
 {}``net fan'' diagram: in the graphs (a) and (b) we have a fan-like
structure of cut Pomerons; in the diagrams (c) and (d)
the cut Pomeron, exchanged between the vertices ($y_{2},\vec b_{2}$)
and  ($y_{3},\vec b_{3}$), is arranged in a zigzag way 
with respect to the  ``fan handle''.\label{cut examples}}
\end{figure}

Before  considering the cuts of the elastic scattering graphs of 
Fig.~\ref{enh-full}, let us apply the AGK cutting procedure to the 
``net fan'' contributions of Fig.~\ref{freve}.
It is convenient to separate various unitarity cuts of {}``net
fan'' graphs in two classes: in the first sub-set cut Pomerons form
 fan-like structures, some examples shown in Fig.~\ref{cut examples}
(a), (b);%
 in the diagrams of the second kind some \textit{cut} Pomerons are connected
to each other in a  zigzag way,
 such that Pomeron end rapidities are arranged
as  $y_{1}>y_{2}<y_3>...$, see Fig.~\ref{cut examples}~(c), (d).

\begin{figure*}[t]
\begin{center}\includegraphics[%
  width=15cm,
  height=3.7cm]{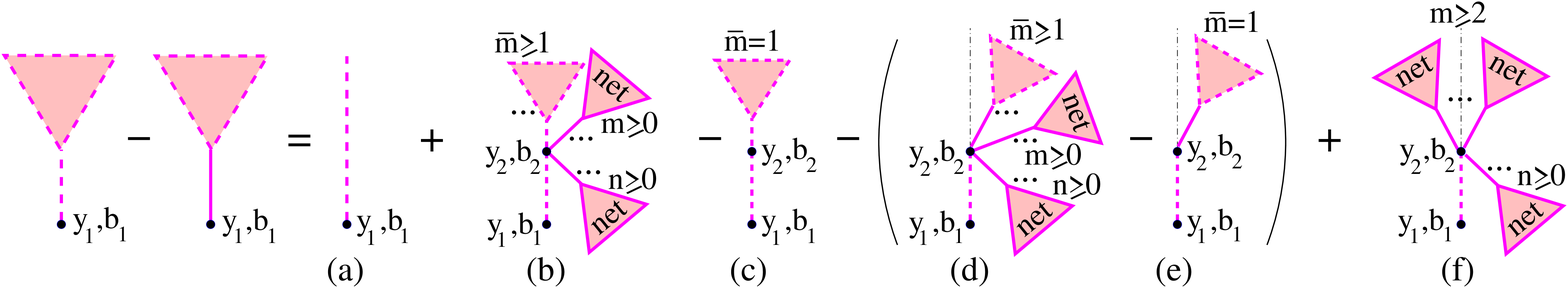}\end{center}
\vspace{-3mm}
\caption{Recursive equation for the contribution 
$2\bar{\chi}_{a(j)|d(l)}^{{\rm fan}}-2\tilde{\chi}_{a(j)|d(l)}^{{\rm fan}}$
of fan-like cuts of {}``net fan'' diagrams, the handle of the fan
being cut. \label{fan-cut-fig}}
\end{figure*}

\begin{figure}[t]
\begin{center}\includegraphics[%
  width=14cm,
  height=4cm]{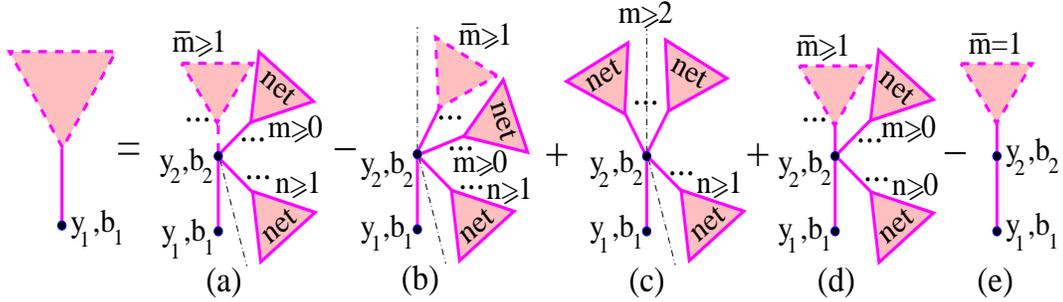}\end{center}
\vspace{-3mm}
\caption{Recursive equation for the contribution $2\tilde{\chi}_{a(j)|d(l)}^{{\rm fan}}$
of fan-like cuts of {}``net fan'' diagrams, the handle
of the fan being uncut.\label{fan-hole-fig}}
\end{figure}

Let us consider the first class and obtain separately both the total
contribution of fan-like cuts $2\bar{\chi}_{a(j)|d(l)}^{{\rm fan}}$
and the part of it, formed by diagrams with the handle of
the fan being uncut, an example shown in Fig.~\ref{cut examples}~(b), 
 $2\tilde{\chi}_{a(j)|d(l)}^{{\rm fan}}$. Applying 
AGK cutting rules to the  graphs of Fig.~\ref{freve}
and collecting contributions of cuts of desirable structures we obtain
for $2\bar{\chi}_{a(j)|d(l)}^{{\rm fan}}-2\tilde{\chi}_{a(j)|d(l)}^{{\rm fan}}$,
$2\tilde{\chi}_{a(j)|d(l)}^{{\rm fan}}$ the representations of 
Figs.~\ref{fan-cut-fig} and \ref{fan-hole-fig},%
 which gives
\begin{eqnarray}
2\bar{\chi}_{a(j)|d(l)}^{{\rm fan}}(y_{1},\vec{b}_{1}|Y,\vec{b})
-2\tilde{\chi}_{a(j)|d(l)}^{{\rm fan}}(y_{1},\vec{b}_{1}|Y,\vec{b})
=2\lambda_{a(j)}\,\chi_{a\mathbb{P}}^{\mathbb{P}}(y_{1},b_{1})\nonumber \\
+G\int_{0}^{y_{1}}\! dy_{2}\int\! d^{2}b_{2}\;
\chi_{\mathbb{PP}}^{\mathbb{P}}
\left\{ \left[(e^{2\bar{\chi}_{a|d}^{{\rm fan}}}-1)\:
e^{-2\chi_{a|d}^{{\rm net}}-2\chi_{d|a}^{{\rm net}}}
-2\bar{\chi}_{a|d}^{{\rm fan}}\right]\right.\nonumber \\
\left.-2\left[(e^{\tilde{\chi}_{a|d}^{{\rm fan}}}-1)\:
e^{-\chi_{a|d}^{{\rm net}}-2\chi_{d|a}^{{\rm net}}}
-\tilde{\chi}_{a|d}^{{\rm fan}}\right]
+(1-e^{-\chi_{a|d}^{{\rm net}}})^{2}\:
e^{-2\chi_{d|a}^{{\rm net}}}\right\} \label{chi-cut-fan}\\
2\tilde{\chi}_{a(j)|d(l)}^{{\rm fan}}(y_{1},\vec{b}_{1}|Y,\vec{b})
=G\int_{0}^{y_{1}}\! dy_{2}\int\! d^{2}b_{2}\;
\chi_{\mathbb{PP}}^{\mathbb{P}}
\left\{ (1-e^{-\chi_{d|a}^{{\rm net}}})\:
e^{-\chi_{d|a}^{{\rm net}}}
\left[(e^{2\bar{\chi}_{a|d}^{{\rm fan}}}-1)\:
e^{-2\chi_{a|d}^{{\rm net}}}\right.\right.\nonumber \\
\left.\left.-2\,(e^{\tilde{\chi}_{a|d}^{{\rm fan}}}-1)\:
e^{-\chi_{a|d}^{{\rm net}}}+(1-e^{-\chi_{a|d}^{{\rm net}}})^{2}\right]
+2\left[(e^{\tilde{\chi}_{a|d}^{{\rm fan}}}-1)\:
e^{-\chi_{a|d}^{{\rm net}}-\chi_{d|a}^{{\rm net}}}
-\tilde{\chi}_{a|d}^{{\rm fan}}\right]\right\} \!.\label{chi-cut-handle}
\end{eqnarray}
Here the omitted indices and arguments of the eikonals in the 
integrands in (\ref{chi-cut-fan}-\ref{chi-cut-handle})
read $\chi_{\mathbb{PP}}^{\mathbb{P}}\equiv
\chi_{\mathbb{PP}}^{\mathbb{P}}(y_{1}-y_{2},|\vec{b}_{1}-\vec{b}_{2}|)$,
 $X_{a|d}\equiv X_{a(j)|d(l)}(y_{2},\vec{b}_{2}|Y,\vec{b})$,
 $X_{d|a}\equiv X_{d(l)|a(j)}(Y-y_{2},\vec{b}-\vec{b}_{2}|Y,\vec{b})$,
$X=\chi^{{\rm net}},\bar{\chi}^{{\rm fan}},\tilde{\chi}^{{\rm fan}}$.

The first diagram in the r.h.s.~of Fig.~\ref{fan-cut-fig} is obtained
by cutting the single Pomeron exchanged between hadron $a$ and the
vertex $(y_{1},b_{1})$ in the r.h.s.~of Fig.~\ref{freve}, whereas
the other ones come from cutting the 2nd graph in the 
r.h.s.~of Fig.~\ref{freve}
in such a way that all cut Pomerons are arranged in a fan-like
structure and the cut plane passes through the handle Pomeron.
 In graph (b) the vertex $(y_{2},b_{2})$ couples together
$\bar{m}\geq1$ cut projectile {}``net fans'', each one characterized
by a fan-like structure of cuts, and any numbers $m,n\geq0$
of uncut projectile and target {}``net fans''. Here one has to subtract the
Pomeron self-coupling contribution ($\bar{m}=1$, $m=n=0$) - graph
(c), as well as the contributions of graphs (d) and (e), where in all
$\bar{m}$ cut projectile {}``net fans'',
 connected to the vertex $(y_{2},b_{2})$,
the handle Pomerons remain uncut and all these
handle Pomerons and all the $m$ uncut projectile {}``net fans''
are situated on the same side of the cut plane. Finally, in graph
(f) the cut plane passes between $m\geq2$ uncut projectile {}``net
fans'', with at least one remained on either side of the cut.

In the recursive representation of Fig.~\ref{fan-hole-fig} for the
contribution $2\tilde{\chi}_{a(j)|d(l)}^{{\rm fan}}$, the graphs (a), (b),
(c) in the r.h.s.~of the Figure are similar to the diagrams (b),
(d), (f) of Fig.~\ref{fan-cut-fig} correspondingly, with the difference
that the handle of the fan is now uncut. Therefore,
there are $n\geq1$  uncut target {}``net fans'' connected to
the vertex $(y_{2},b_{2})$ in  such a way that at least one of them
is positioned on the opposite side of the cut plane with respect to
the handle Pomeron. On the other hand, one has to add graph
(d), where the vertex $(y_{2},b_{2})$ couples together $\bar{m}\geq1$
 projectile {}``net fans'', which are cut in a fan-like
way and have their handle Pomerons uncut and positioned on the same
side of the cut plane, together with any numbers $m\geq0$ of projectile
and $n\geq0$ of target uncut {}``net fans'', such that the vertex
remains uncut. Here one has to subtract  the Pomeron self-coupling
($\bar{m}=1$, $m=n=0$) -- graph (e).

Adding (\ref{chi-cut-handle}) to (\ref{chi-cut-fan}), we obtain
\begin{eqnarray}
2\bar{\chi}_{a(j)|d(l)}^{{\rm fan}}(y_{1},\vec{b}_{1}|Y,\vec{b})
=2\lambda_{a(j)}\,\chi_{a\mathbb{P}}^{\mathbb{P}}(y_{1},b_{1}) \nonumber \\
+G\int_{0}^{y_{1}}\!\! dy_{2}\int\!\! d^{2}b_{2}\;
\chi_{\mathbb{PP}}^{\mathbb{P}}\,
 \left\{ \left[(e^{2\bar{\chi}_{a|d}^{{\rm fan}}}-1)
\,e^{-2\chi_{a|d}^{{\rm net}}}
+(1-e^{-\chi_{a|d}^{{\rm net}}})^{2}\right]e^{-\chi_{d|a}^{{\rm net}}}
-2\bar{\chi}_{a|d}^{{\rm fan}}\right\} \!,\!\!\label{chi-cut-fan1}
\end{eqnarray}
with the solution (c.f.~(\ref{net-fan}))
\begin{eqnarray}
\bar{\chi}_{a(j)|d(l)}^{{\rm fan}}(y_{1},\vec{b}_{1}|Y,\vec{b})=
\chi_{a(j)|d(l)}^{{\rm net}}(y_{1},\vec{b}_{1}|Y,\vec{b}).\label{equiv}
\end{eqnarray}

\begin{figure*}[t]
\begin{center}\includegraphics[%
  width=12cm,
  height=6cm]{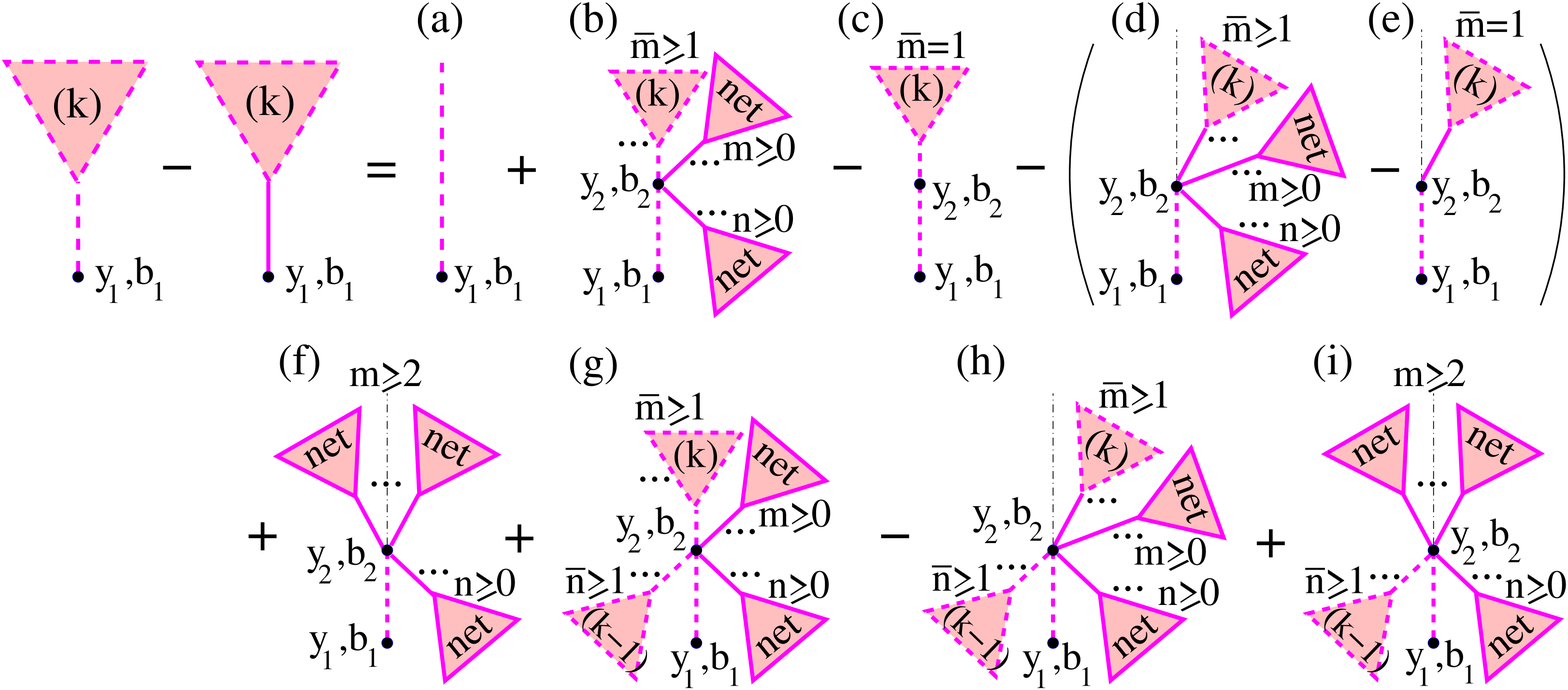}\end{center}
\vspace{-4mm}
\caption{Recursive equation for the  $k$-th order  cut {}``net-fan'' contribution 
$2\bar{\chi}_{a(j)|d(l)}^{\rm{net}(k)}
-2\tilde{\chi}_{a(j)|d(l)}^{\rm{net}(k)}$ with cut ``handle''.
\label{netcut-fig}}
\end{figure*}

\begin{figure*}[t]
\begin{center}\includegraphics[%
  width=15cm,
  height=6cm]{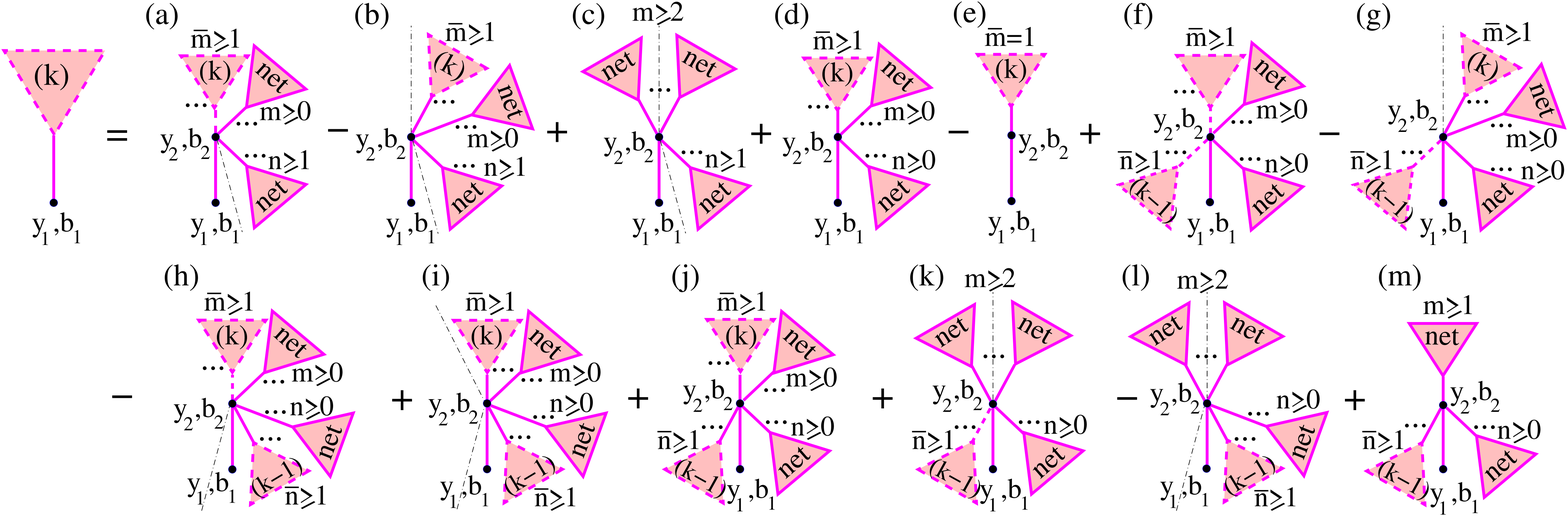}\end{center}
\vspace{-4mm}
\caption{Recursive equation for the $k$-th order cut {}``net-fan'' contribution 
$2\tilde{\chi}_{a(j)|d(l)}^{\rm{net}(k)}$ with uncut ``handle''.
 \label{netcuth-fig}}
\end{figure*}

To investigate  zigzag-like cuts of ``net fan'' graphs,
the examples shown in Fig.~\ref{cut examples}~(c) and (d), we introduce
 $k$-th order cut {}``net fan'' contributions $2\bar{\chi}_{a(j)|d(l)}^{{\rm net}
(k)}$, $k\geq 2$, which in addition to the above-considered 
fan-like cut diagrams contain also ones with
 up to $k$  \textit{cut} Pomerons connected to each other in a 
zigzag way, i.e., with Pomeron end rapidities being arranged as
$y_{1}>y_{2}<...>y_{k+1}$. For example, the graphs of 
 Fig.~\ref{cut examples}~(c) and (d) belong correspondingly 
to the 2nd and 3rd order cut ``net fan''  contributions. 
As before, we consider   two subsamples
of the  diagrams, with the handle Pomerons being  cut, 
$2\bar{\chi}_{a(j)|d(l)}^{{\rm net}(k)}
-2\tilde{\chi}_{a(j)|d(l)}^{{\rm net}(k)}$,
and uncut, $2\tilde{\chi}_{a(j)|d(l)}^{{\rm net}(k)}$, which leads us to the
recursive equations of Figs.~\ref{netcut-fig} and \ref{netcuth-fig}
respectively.%
 Compared to the ones of Figs.~\ref{fan-cut-fig} and \ref{fan-hole-fig},
they contain additional graphs, Fig.~\ref{netcut-fig}~(g)-(i) and
Fig.~\ref{netcuth-fig}~(f)-(m), where the vertex $(y_{2},\vec{b}_{2})$
is coupled to $\bar{n}\geq1$ cut target {}``net fans'' of $(k-1)$-th order 
(we set $\bar{\chi}_{a(j)|d(l)}^{{\rm net}(1)}\equiv\bar{\chi}_{a(j)|d(l)}^{{\rm fan}}$,
$\tilde{\chi}_{a(j)|d(l)}^{{\rm net}(1)}\equiv\tilde{\chi}_{a(j)|d(l)}^{{\rm fan}}$).
Thus, we obtain
\begin{eqnarray}
2\bar{\chi}_{a(j)|d(l)}^{{\rm net}(k)}(y_{1},\vec{b}_{1}|Y,\vec{b})
-2\tilde{\chi}_{a(j)|d(l)}^{{\rm net}(k)}(y_{1},\vec{b}_{1}|Y,\vec{b})
=2\lambda_{a(j)}\,\chi_{a\mathbb{P}}^{\mathbb{P}}(y_{1},b_{1})\nonumber \\
+G\int_{0}^{y_{1}}\! dy_{2}\int\! d^{2}b_{2}\;
\chi_{\mathbb{PP}}^{\mathbb{P}}\,
\left\{ \left[(e^{2\bar{\chi}_{a|d}^{{\rm net}(k)}}-1)\:
e^{-2\chi_{a|d}^{{\rm net}}-2\chi_{d|a}^{{\rm net}}}
-2\bar{\chi}_{a|d}^{{\rm net}(k)}\right]\right.\nonumber \\
-2\left[(e^{\tilde{\chi}_{a|d}^{{\rm net}(k)}}-1)\:
e^{-\chi_{a|d}^{{\rm net}}-2\chi_{d|a}^{{\rm net}}}
-\tilde{\chi}_{a|d}^{{\rm net}(k)}\right]
+(1-e^{-\chi_{a|d}^{{\rm net}}})^{2}\:e^{-2\chi_{d|a}^{{\rm net}}}\nonumber \\
\left.+\left[(e^{2\bar{\chi}_{a|d}^{{\rm net}(k)}}-1)\:e^{-2\chi_{a|d}^{{\rm net}}}
-2(e^{\tilde{\chi}_{a|d}^{{\rm net}(k)}}-1)\:
e^{-\chi_{a|d}^{{\rm net}}}+(1-e^{-\chi_{a|d}^{{\rm net}}})^{2}\right]\,
 (e^{2\bar{\chi}_{d|a}^{{\rm net}(k-1)}}-1)\:e^{-2\chi_{d|a}^{{\rm net}}}
\right\} \label{chi-cut-(k)}\\
2\tilde{\chi}_{a(j)|d(l)}^{{\rm net}(k)}(y_{1},\vec{b}_{1}|Y,\vec{b})
=G\int_{0}^{y_{1}}\! dy_{2}\int\! d^{2}b_{2}\;
\chi_{\mathbb{PP}}^{\mathbb{P}}
\left\{ (1-e^{-\chi_{d|a}^{{\rm net}}})\:e^{-\chi_{d|a}^{{\rm net}}}
\left[(e^{2\bar{\chi}_{a|d}^{{\rm net}(k)}}-1)\:
e^{-2\chi_{a|d}^{{\rm net}}}\right.\right.\nonumber \\
-\left.2\,(e^{\tilde{\chi}_{a|d}^{{\rm net}(k)}}-1)\:
e^{-\chi_{a|d}^{{\rm net}}}+(1-e^{-\chi_{a|d}^{{\rm net}}})^{2}\right]
+2\left[(e^{\tilde{\chi}_{a|d}^{{\rm net}(k)}}-1)\:
e^{-\chi_{a|d}^{{\rm net}}-\chi_{d|a}^{{\rm net}}}
-\tilde{\chi}_{a|d}^{{\rm net}(k)}\right]\nonumber \\
 -\left[(e^{2\bar{\chi}_{a|d}^{{\rm net}(k)}}-1)\:e^{-2\chi_{a|d}^{{\rm net}}}
-2(e^{\tilde{\chi}_{a|d}^{{\rm net}(k)}}-1)\:
e^{-\chi_{a|d}^{{\rm net}}}+(1-e^{-\chi_{a|d}^{{\rm net}}})^{2}\right]\,
 (e^{2\bar{\chi}_{d|a}^{{\rm net}(k-1)}}-1)\:e^{-2\chi_{d|a}^{{\rm net}}}\nonumber \\
 +\left. 2\left[(e^{2\bar{\chi}_{a|d}^{{\rm net}(k)}}-1)\,
e^{-2\chi_{a|d}^{{\rm net}}}+(1-e^{-\chi_{a|d}^{{\rm net}}})^{2}
-2\,(1-e^{-\chi_{a|d}^{{\rm net}}})\right]\,
(e^{\tilde{\chi}_{d|a}^{{\rm net}(k-1)}}-1)\:e^{-\chi_{d|a}^{{\rm net}}}
\right\}  \!.\label{chi-cut-handle(k)}
\end{eqnarray}

Adding (\ref{chi-cut-(k)}) to (\ref{chi-cut-handle(k)}), we obtain 
\begin{eqnarray}
2\bar{\chi}_{a(j)|d(l)}^{{\rm net}(k)}(y_{1},\vec{b}_{1}|Y,\vec{b})
=2\lambda_{a(j)}\,\chi_{a\mathbb{P}}^{\mathbb{P}}(y_{1},b_{1})\nonumber \\
+G\int_{0}^{y_{1}}\!\! dy_{2}\int\!\! d^{2}b_{2}\;
\chi_{\mathbb{PP}}^{\mathbb{P}}\,
 \left\{\left[(e^{2\bar{\chi}_{a|d}^{{\rm net}(k)}}-1)
\,e^{-2\chi_{a|d}^{{\rm net}}}+
(1-e^{-\chi_{a|d}^{{\rm net}}})^{2}\right]
e^{-\chi_{d|a}^{{\rm net}}}-2\bar{\chi}_{a|d}^{{\rm net}(k)}\right.\nonumber \\
+\left. 2\left[(e^{2\bar{\chi}_{a|d}^{{\rm net}(k)}}-1)\,
e^{-2\chi_{a|d}^{{\rm net}}}+(1-e^{-\chi_{a|d}^{{\rm net}}})^{2}
-2\,(1-e^{-\chi_{a|d}^{{\rm net}}})\right]\,
(e^{\tilde{\chi}_{d|a}^{{\rm net}(k-1)}}-1)\:e^{-\chi_{d|a}^{{\rm net}}}
\right\} \!,\label{chi-net-(k)}
\end{eqnarray}
with the solution (c.f.~(\ref{net-fan}))
\begin{eqnarray}
\bar{\chi}_{a(j)|d(l)}^{{\rm net}(k)}(y_{1},\vec{b}_{1}|Y,\vec{b})=
\chi_{a(j)|d(l)}^{{\rm net}}(y_{1},\vec{b}_{1}|Y,\vec{b})\,.\label{equiv1}
\end{eqnarray}
Thus, for the summary contribution of all cuts of {}``net fan''
graphs of Fig.~\ref{enh-full} we obtain 
$2\bar{\chi}_{a(j)|d(l)}^{{\rm net}}\equiv\lim_{k\rightarrow\infty}
2\bar{\chi}_{a(j)|d(l)}^{{\rm net}(k)}
=\chi_{a(j)|d(l)}^{{\rm net}}$,
as it should be. On the other hand, contributions of various zigzag-like
cuts precisely cancel  each other
\begin{eqnarray}
2\bar{\chi}_{a(j)|d(l)}^{{\rm zz}(k)}
\equiv 2\bar{\chi}_{a(j)|d(l)}^{{\rm net}(k)}-2\bar{\chi}_{a|d}^{{\rm net}(k-1)}= 0
\label{zz=0} \\
2\bar{\chi}_{a(j)|d(l)}^{{\rm zz}}\equiv 
2\bar{\chi}_{a(j)|d(l)}^{{\rm net}}-2\bar{\chi}_{a(j)|d(l)}^{{\rm fan}}
= \sum_{k=2}^{\infty}2\bar{\chi}_{a(j)|d(l)}^{{\rm zz}(k)}= 0.
\end{eqnarray}

Making use of (\ref{equiv1}), we can re-write   (\ref{chi-cut-(k)}) as
\begin{eqnarray}
2\left[\bar{\chi}_{a(j)|d(l)}^{{\rm net}(k)}(y_{1},\vec{b}_{1}|Y,\vec{b})
-\tilde{\chi}_{a(j)|d(l)}^{{\rm net}(k)}(y_{1},\vec{b}_{1}|Y,\vec{b})\right]
=2\lambda_{a(j)}\,\chi_{a\mathbb{P}}^{\mathbb{P}}(y_{1},b_{1})\nonumber \\
+2G\int_{0}^{y_{1}}\! dy_{2}\int\! d^{2}b_{2}\;
\chi_{\mathbb{PP}}^{\mathbb{P}}\,
\left\{1- e^{-\left[\bar{\chi}_{a|d}^{{\rm net}(k)}
-\tilde{\chi}_{a|d}^{{\rm net}(k)}\right]}
-\left[\bar{\chi}_{a|d}^{{\rm net}(k)}-\tilde{\chi}_{a|d}^{{\rm net}(k)}\right]
\right\} \label{newfan}
\end{eqnarray}
and obtain (c.f.~(\ref{fan}))
\begin{eqnarray}
\bar{\chi}_{a(j)|d(l)}^{{\rm net}(k)}(y_{1},\vec{b}_{1}|Y,\vec{b})
-\tilde{\chi}_{a(j)|d(l)}^{{\rm net}(k)}(y_{1},\vec{b}_{1}|Y,\vec{b})
=\chi_{a(j)}^{{\rm fan}}(y_{1},{b}_{1})\,,\;\;\;\;k\geq 2,
\end{eqnarray}
i.e.~the summary contribution of all AGK cuts of {}``net fan'' graphs, with the
cut plane passing through the handle Pomeron, satisfies the usual fan diagram
equation  (\ref{fan}), being independent on re-scatterings on the partner hadron.

\begin{figure*}[t]
\begin{center}\includegraphics[%
  width=15cm,
  height=3cm]{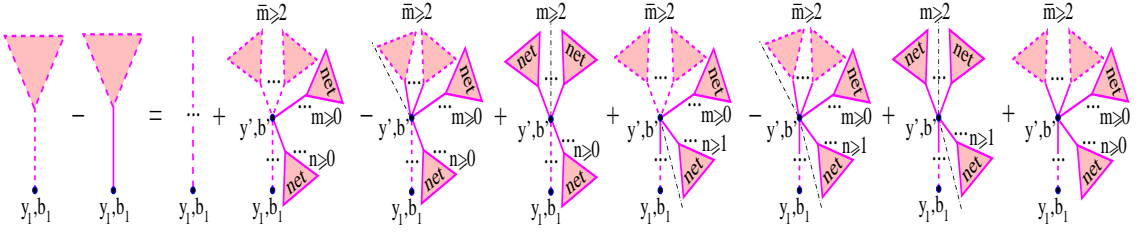}\end{center}
\vspace{-3mm}
\caption{Alternative representation for the fan-like cut contribution 
$2\bar{\chi}_{a(j)|d(l)}^{{\rm fan}}-2\tilde{\chi}_{a(j)|d(l)}^{{\rm fan}}$,
with the handle Pomeron being cut. Each broken Pomeron line
denotes a $t$-channel sequence of Pomerons which are separated by 
 vertices connected to uncut
projectile and target {}``net fans''.\label{fancut2-fig}}
\end{figure*}

\begin{figure*}[t]
\begin{center}\includegraphics[%
  width=15cm,
  height=3cm]{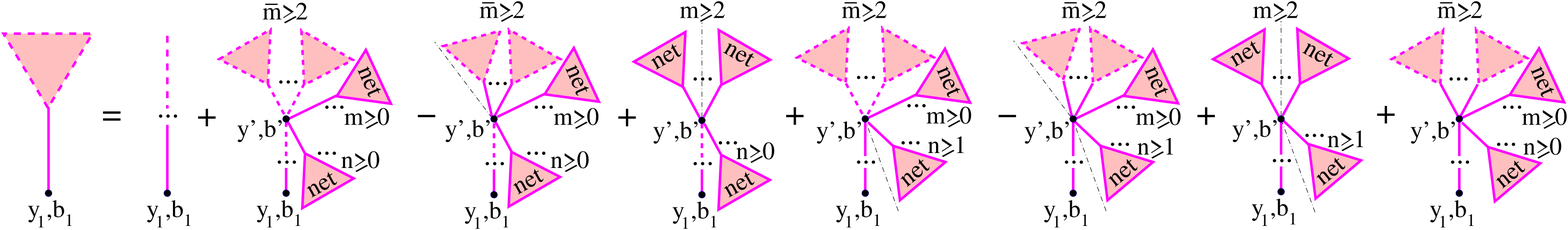}\end{center}
\vspace{-3mm}
\caption{Alternative representation for the fan-like cut  contribution 
$2\tilde{\chi}_{a(j)|d(l)}^{{\rm fan}}$,  with the handle Pomeron being uncut. 
The broken Pomeron lines have the same meaning as in
Fig.~\ref{fancut2-fig}. \label{fancuth2-fig}}
\end{figure*}

\begin{figure}[t]
\begin{center}\includegraphics[%
  width=11cm,
  height=11.cm]{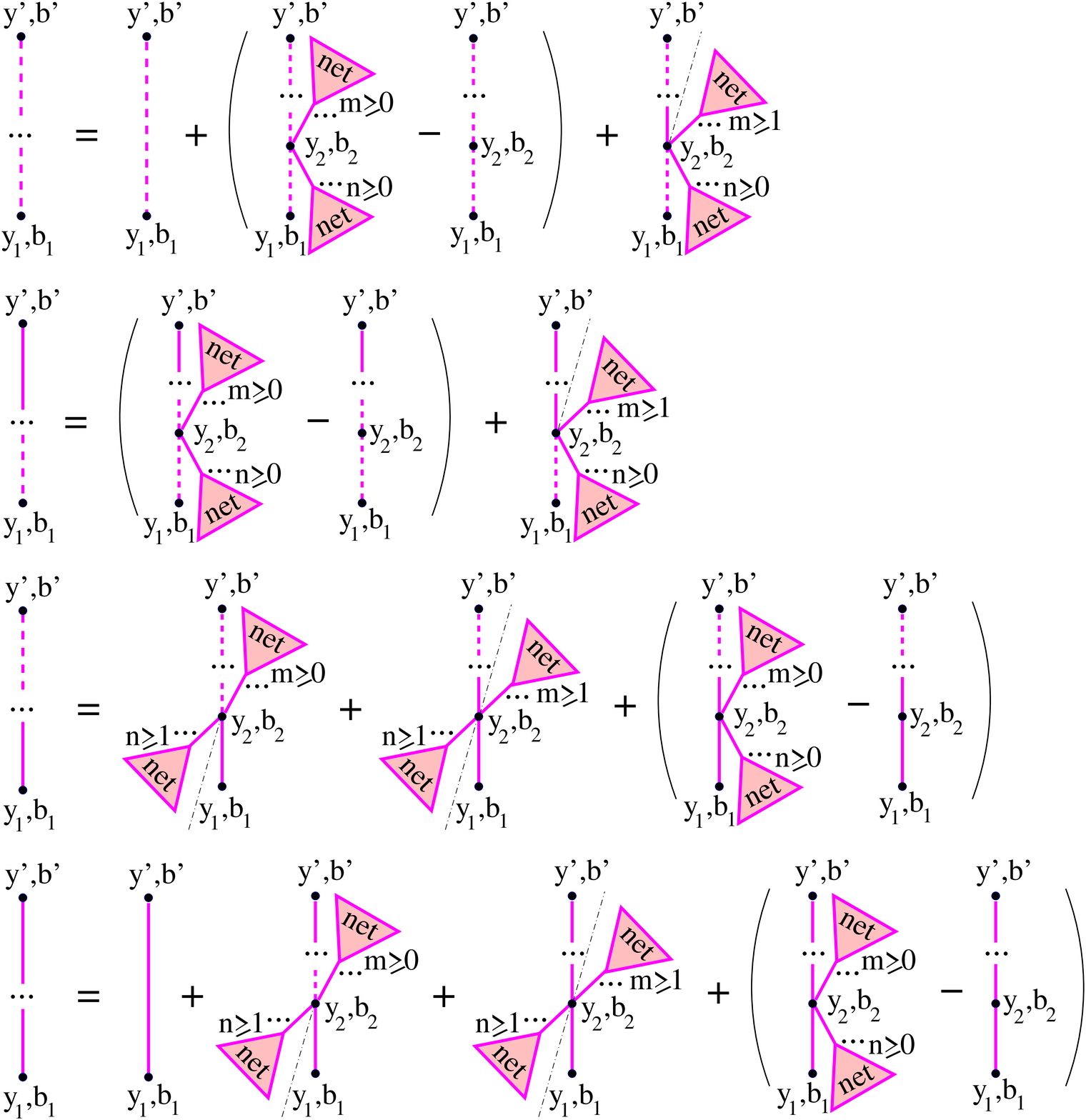}\end{center}
\vspace{-3mm}
\caption{Recursive equations in the Figure generate  
$t$-channel sequences of cut and uncut Pomerons,
 which are separated by multi-Pomeron vertices,
 each one being connected to at least one uncut 
 projectile or target  {}``net fan''.\label{broke-pom}}
\end{figure}

One can obtain an alternative representation for $\bar{\chi}_{a(j)|d(l)}^{{\rm fan}}$,
$\tilde{\chi}_{a(j)|d(l)}^{{\rm fan}}$, as shown in
Figs.~\ref{fancut2-fig} and \ref{fancuth2-fig},%
 applying (\ref{chi-cut-fan}-\ref{chi-cut-handle}) (correspondingly 
 Figs.~\ref{fan-cut-fig} and \ref{fan-hole-fig})
recursively to generate any number of vertices, connected to \textsl{uncut}
projectile and target {}``net fans'', along the handle of the fan.
 The broken Pomeron lines in  Figs.~\ref{fancut2-fig} and \ref{fancuth2-fig} 
  correspond to  $t$-channel sequences of cut and uncut Pomerons,
 which are separated by  vertices connected to uncut projectile and target
  {}``net fans''; the corresponding contributions  are defined via recursive
  representations of Fig.~\ref{broke-pom}.%
In particular, the contributions $2\chi_{a(j)|d(l)}^{\mathbb{P}_{{\rm cc}}}$ and
$2\chi_{a(j)|d(l)}^{\mathbb{P}_{{\rm uc}}}$ of the first graphs
in the r.h.s.~of Figs.~\ref{fancut2-fig} and \ref{fancuth2-fig}
respectively (the index ${\mathbb{P}_{xy}}$ indicates whether the
downmost (uppermost) Pomeron in the sequence is cut, 
$x={\rm c}$ ($y={\rm c}$),  or uncut, $x={\rm u}$ ($y={\rm u}$))
are defined as (c.f.~(\ref{chi-cut-fan}-\ref{chi-cut-handle}))
\begin{eqnarray}
2\chi_{a(j)|d(l)}^{\mathbb{P}_{{\rm cc}}}(y_{1},\vec b_{1}|Y,\vec{b})
=2\lambda_{a(j)}\,\chi_{a\mathbb{P}}^{\mathbb{P}}(y_{1},b_{1})
+2G\int_{0}^{y_{1}}\! dy_2 \int\! d^{2}b_2 
\;\chi_{\mathbb{PP}}^{\mathbb{P}}\nonumber \\
\times \left[
(e^{-2\chi_{a|d}^{{\rm net}}-2\chi_{d|a}^{{\rm net}}}-1)\,
(\chi_{a|d}^{\mathbb{P}_{{\rm cc}}}+
\chi_{a|d}^{\mathbb{P}_{{\rm uc}}})
- (e^{-\chi_{a|d}^{{\rm net}}-2\chi_{d|a}^{{\rm net}}}-1) \,
\chi_{a|d}^{\mathbb{P}_{{\rm uc}}}\right]  \label{chi-a-cc}\\
2\chi_{a(j)|d(l)}^{\mathbb{P}_{{\rm uc}}}(y_{1},\vec b_{1}|Y,\vec{b})
=2G\int_{0}^{y_{1}}\! dy_2\int\! d^{2}b_2
\;\chi_{\mathbb{PP}}^{\mathbb{P}}\nonumber \\
\times\left[(1-e^{-\chi_{d|a}^{{\rm net}}})\,
e^{-2\chi_{a|d}^{{\rm net}}-\chi_{d|a}^{{\rm net}}}
 \,(\chi_{a|d}^{\mathbb{P}_{{\rm cc}}}+
\chi_{a|d}^{\mathbb{P}_{{\rm uc}}}) 
+(e^{-\chi_{a|d}^{{\rm net}}-2\chi_{d|a}^{{\rm net}}}-1)\,
\chi_{a|d}^{\mathbb{P}_{{\rm uc}}}\right] \!.\label{chi-a-uc}
\end{eqnarray}

\begin{figure*}[t]
\begin{center}\includegraphics[%
  width=15cm,
  height=9cm]{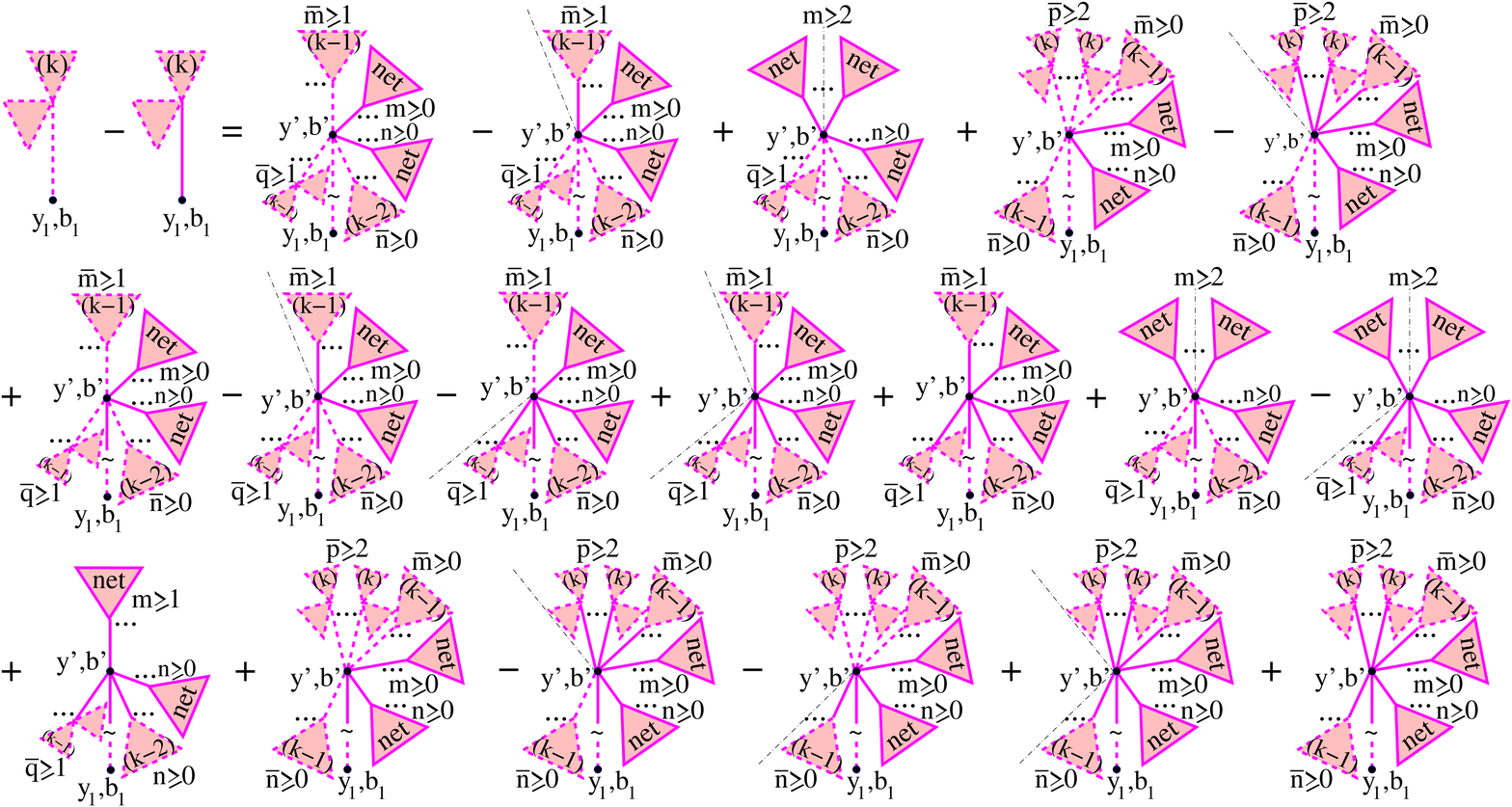}\end{center}
\vspace{-3mm}
\caption{Recursive equation for the contribution 
$2\bar{\chi}_{a(j)|d(l)}^{{\rm zz}(k)}-2\tilde{\chi}_{a(j)|d(l)}^{{\rm zz}(k)}$
of zigzag-like cuts of {}``net-fan'' diagrams,  the handle Pomeron being cut.
The broken Pomeron lines  between the vertices $(y_1, \vec{b}_1)$ and 
$(y',\vec b')$ correspond  here to  $t$-channel sequences of cut and uncut Pomerons,
 which are separated by  multi-Pomeron vertices   connected to uncut
and $(k-1)$-th order cut projectile and target {}``net fans''.\label{zigcut2-fig}}
\end{figure*}

Similarly, we can  obtain a recursive representation 
for $k$-th order zigzag-like cut {}``net fans''
 $2\bar{\chi}_{a(j)|d(l)}^{{\rm zz}(k)}
\equiv2\bar{\chi}_{a(j)|d(l)}^{{\rm net}(k)}-2\bar{\chi}_{a(j)|d(l)}^{{\rm net}(k-1)}$, 
$2\tilde{\chi}_{a(j)|d(l)}^{{\rm zz}(k)}
\equiv 2\tilde{\chi}_{a(j)|d(l)}^{{\rm net}(k)}-2\tilde{\chi}_{a(j)|d(l)}^{{\rm net}(k-1)}$, 
 applying recursively the relations (\ref{chi-cut-(k)}-\ref{chi-cut-handle(k)})
  (Figs.~\ref{netcut-fig} and \ref{netcuth-fig}) to generate any number
of intermediate vertices along the handle Pomeron, which are connected
to uncut and $(k-1)$-th order cut projectile and target 
 {}``net fans'', until we end up with  the vertex $(y',\vec{b}')$, which either
 couples together $\bar p\geq 2$ $k$-th order projectile zigzag-like cut
contributions and any numbers of uncut and  $(k-1)$-th order cut projectile 
and target {}``net fans'',  or is coupled to $\bar q\geq 1$
$(k-1)$-th order target  zigzag-like cut contributions
(in addition, to any numbers of uncut and  $(k-2)$-th order cut  
 target {}``net fans'') and to uncut and  
$(k-1)$-th order cut projectile  {}``net fans''.
The corresponding relation for $2\bar{\chi}_{a(j)|d(l)}^{{\rm zz}(k)}- 
2\tilde{\chi}_{a(j)|d(l)}^{{\rm zz}(k)}$ is shown  in Fig.~\ref{zigcut2-fig},%
the one for $2\tilde{\chi}_{a(j)|d(l)}^{{\rm zz}(k)}$ looks similarly
(c.f.~Figs.~\ref{fancut2-fig} and \ref{fancuth2-fig}).

\section{Cut enhanced diagrams\label{enh.sec} }

We are going to derive the complete set of cut diagrams corresponding
to $s$-channel discontinuity of elastic scattering contributions of 
Fig.~\ref{enh-full}.
Let us start with cut graphs characterized by a tree-like
structure of cut Pomerons, which can be constructed coupling any numbers
$\bar{m},\bar{n}$ of fan-like cut projectile and 
target {}``net fans'' in one vertex. 

\begin{figure*}[t]
\begin{center}\includegraphics[%
  width=15cm,
  height=7cm]{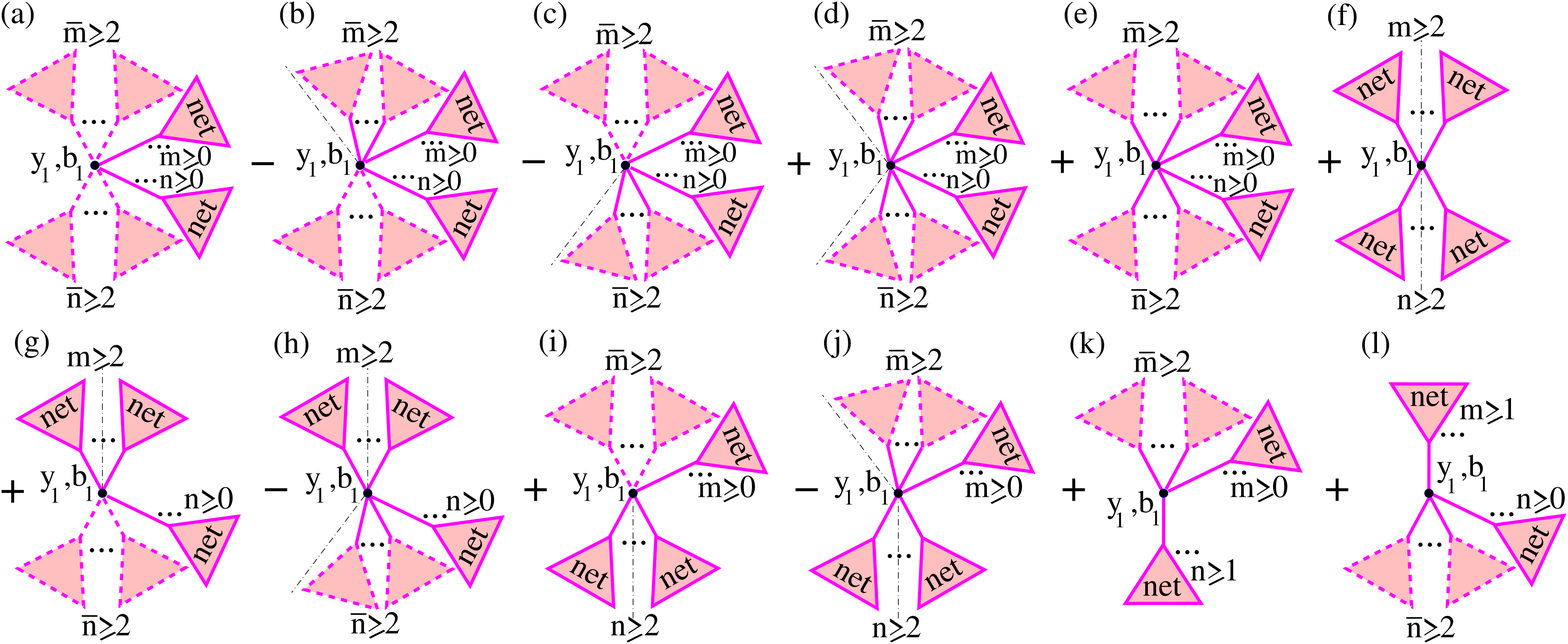}\end{center}
\vspace{-3mm}
\caption{Tree-like cut enhanced diagrams. 
The vertex  $(y_{1},\vec{b}_{1})$ couples together $\bar{m}$ projectile
 and $\bar{n}$ target fan-like cut
{}``net fans''; $\bar{m},\bar{n}\neq 1$.\label{treecut1-fig}}
\end{figure*}

First we consider the case of $\bar{m},\bar{n}\neq1$, which leads us to the
set of graphs of Fig.~\ref{treecut1-fig},%
 where we do not have any double counting
of the same contributions. For example, the graphs Fig.~\ref{treecut1-fig}~(a)-(e)
 have a single vertex $(y_{1},\vec{b}_{1})$, which couples
together $\bar{m}\geq2$ projectile and $\bar{n}\geq2$ target 
 fan-like cut  {}``net fans''.
 Correspondingly, different structures of cut  {}``net fans''
and different topologies of the uncut ones  result in
different diagrams. For the combined contribution of all the graphs
of Fig.~\ref{treecut1-fig} we obtain, using (\ref{equiv}),
\begin{eqnarray}
2\bar{\chi}_{ad(jl)}^{{\rm tree(1)}}(Y,b)
=2G\int_{0}^{Y}\!dy_{1}\int\! d^{2}b_{1} \left\{(1-e^{-\chi_{a|d}^{{\rm net}}}
 -\chi_{a|d}^{{\rm net}}\: e^{-2\chi_{a|d}^{{\rm net}}})\;
(1-e^{-\chi_{d|a}^{{\rm net}}}-\chi_{d|a}^{{\rm net}}
\: e^{-2\chi_{d|a}^{{\rm net}}} )\right. \nonumber \\
+\left. \chi_{a|d}^{{\rm net}}\;
e^{-2\chi_{a|d}^{{\rm net}}-\chi_{d|a}^{{\rm net}}}\;
(e^{\tilde \chi_{d|a}^{{\rm fan}}}-1-\tilde \chi_{d|a}^{{\rm fan}})
+ \chi_{d|a}^{{\rm net}}\;
e^{-\chi_{a|d}^{{\rm net}}-2\chi_{d|a}^{{\rm net}}}\;
(e^{\tilde \chi_{a|d}^{{\rm fan}}}-1-\tilde \chi_{a|d}^{{\rm fan}})
\right\}\!,
\label{chi-tree-1}
\end{eqnarray}
where we use the abbreviations 
$X_{a|d}\equiv X_{a(j)|d(l)}(Y-y_{1},\vec{b}-\vec{b}_{1}|Y,\vec{b})$,
$X_{d|a}\equiv X_{d(l)|a(j)}(y_{1},\vec{b}_{1}|Y,\vec{b})$,
$X=\chi^{{\rm net}}$, $\tilde \chi^{{\rm fan}}$.

\begin{figure*}[t]
\begin{center}\includegraphics[%
  width=15cm,
  height=3.5cm]{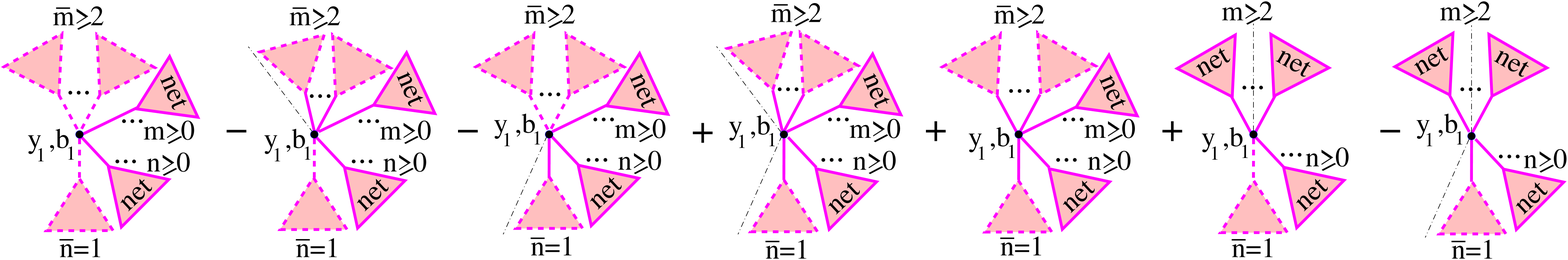}\end{center}
\vspace{-3mm}
\caption{The same as in Fig.~\ref{treecut1-fig} for $\bar{m}\neq1$  and $\bar{n}=1$.
\label{treecut2-fig}}
\end{figure*}

Now we come to the case of $\bar{m}\neq1$  and $\bar{n}=1$,
which results in the diagrams of Fig.~\ref{treecut2-fig}.%
 Similarly to the above, we obtain
\begin{eqnarray}
2\bar{\chi}_{ad(jl)}^{{\rm tree(2)}}(Y,b)=2G
\int_{0}^{Y}\! dy_{1}\int\! d^{2}b_{1}
 \left\{(1-e^{-\chi_{a|d}^{{\rm net}}}
 - \chi_{a|d}^{{\rm net}}\: e^{-2\chi_{a|d}^{{\rm net}}})\;
(\chi_{d|a}^{{\rm net}}\: e^{-2\chi_{d|a}^{{\rm net}}}
- \tilde \chi_{d|a}^{{\rm fan}}\:e^{-\chi_{d|a}^{{\rm net}}} )
\right.\nonumber \\
\left.- \chi_{d|a}^{{\rm net}}\;
e^{-\chi_{a|d}^{{\rm net}}-2\chi_{d|a}^{{\rm net}}}\;
(e^{\tilde \chi_{a|d}^{{\rm fan}}}-1-\tilde \chi_{a|d}^{{\rm fan}})
\right\}\!.
\label{chi-tree-2}
\end{eqnarray}

\begin{figure}[t]
\begin{center}\includegraphics[%
  width=15cm,
  height=4.5cm]{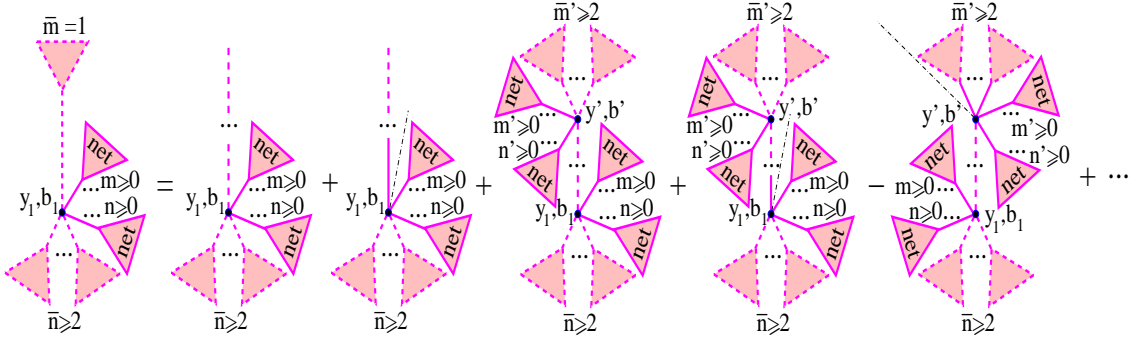}\end{center}
\vspace{-3mm}
\caption{Cut diagram on the l.h.s.~can be expanded as shown in the picture.\label{treeexp-fig}}
\end{figure}

\begin{figure*}[t]
\begin{center}\includegraphics[%
  width=15cm,
  height=3cm]{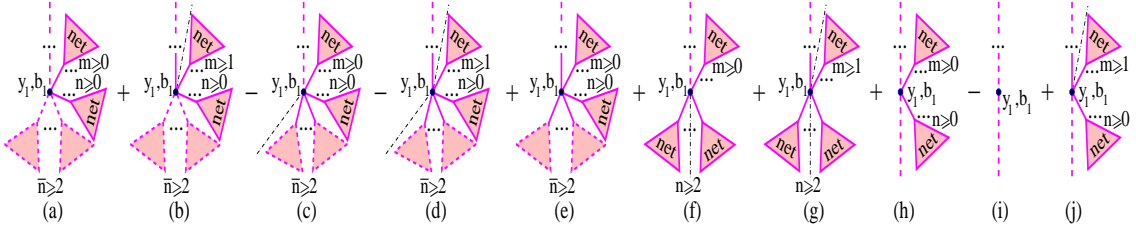}\end{center}
\vspace{-3mm}
\caption{Additional tree-like cut  diagrams, not included in
Figs.~\ref{treecut1-fig}, \ref{treecut2-fig}.\label{treecut3-fig}}
\end{figure*}

 Finally we consider the case of $\bar{n}\neq1$ and $\bar{m}=1$,
which can be obtained reversing the graphs of Fig.~\ref{treecut2-fig}
upside-down. There we have to correct for double counting of
the same contributions. For example, considering the first diagram
of Fig.~\ref{treecut2-fig} 
 being reversed upside-down and expanding its projectile fan-like
cut {}``net fan'' using the relations of Figs.~\ref{fancut2-fig} and
\ref{fancuth2-fig}, we obtain the set of graphs of Fig.~\ref{treeexp-fig}.%
  Clearly, the third diagram in the r.h.s.~of
Fig.~\ref{treeexp-fig}, being symmetric with respect to the projectile and the target,
will appear in a similar expansion of  the first graph of Fig.~\ref{treecut2-fig}.
On the other hand, all the other graphs in the r.h.s.~of Fig.~\ref{treeexp-fig},
except the first two, find their duplicates in the expansions of other
diagrams of Fig.~\ref{treecut2-fig}. Thus, the only new contributions are the ones 
of Fig.~\ref{treecut3-fig}~(a)-(g).%
 In addition, we have to include the graphs
(h)-(j) of the Figure, which correspond  to a $t$-channel sequence
 of $l\geq 2$ cut and uncut Pomerons which are separated by 
 vertices connected to uncut projectile and target {}``net fans'',
with the downmost and the uppermost Pomerons in the sequence being cut.
 The  contribution of the graphs of  Fig.~\ref{treecut3-fig} is
\begin{eqnarray}
2\bar{\chi}_{ad(jl)}^{{\rm tree(3)}}\!(Y,b)=2G
\int_{0}^{Y}\!dy_{1}\int\! d^{2}b_{1}
 \left\{\left[(\chi_{a|d}^{\mathbb{P}_{{\rm cc}}}
+\chi_{a|d}^{\mathbb{P}_{{\rm uc}}})\;
 e^{-2\chi_{a|d}^{{\rm net}}}
-\chi_{a|d}^{\mathbb{P}_{{\rm uc}}}\; e^{-\chi_{a|d}^{{\rm net}}}\right]
\right.\nonumber \\
\times \;(1- e^{-\chi_{d|a}^{{\rm net}}} - \chi_{d|a}^{{\rm net}}
\: e^{-2\chi_{d|a}^{{\rm net}}}) 
-(\chi_{a|d}^{\mathbb{P}_{{\rm cc}}}+
\chi_{a|d}^{\mathbb{P}_{{\rm uc}}})\;
(e^{\tilde \chi_{d|a}^{{\rm fan}}}-1-\tilde \chi_{d|a}^{{\rm fan}})
 \: e^{-2\chi_{a|d}^{{\rm net}}-\chi_{d|a}^{{\rm net}}}\nonumber \\
+\left.\lambda_{d(l)}\,\chi_{d\mathbb{P}}^{\mathbb{P}}(y_{1},b_{1})
\left[\chi_{a|d}^{\mathbb{P}_{{\rm cc}}} \:
(e^{-2\chi_{a|d}^{{\rm net}}-2\chi_{d|a}^{{\rm net}}}-1)
-\chi_{a|d}^{\mathbb{P}_{{\rm uc}}}\:
e^{-\chi_{a|d}^{{\rm net}}-2\chi_{d|a}^{{\rm net}}}\:
(1-e^{-\chi_{a|d}^{{\rm net}}})\right]\right\}\!.
\label{chi-tree-3}
\end{eqnarray}

Adding (\ref{chi-tree-1}-\ref{chi-tree-3}) together and using (\ref{net-fan}),
 (\ref{chi-cut-handle}),  (\ref{equiv}), 
(\ref{chi-a-cc}-\ref{chi-a-uc}),  we can obtain
\begin{eqnarray}
2\bar{\chi}_{ad(jl)}^{{\rm tree}}(Y,b)=\sum _{i=1}^3 
2\bar{\chi}_{ad(jl)}^{{\rm tree(i)}}(Y,b)
=2G\int_{0}^{Y}\! dy_{1}
 \int\! d^{2}b_{1}\left\{ (1-e^{-\chi_{a|d}^{{\rm net}}})\,
(1-e^{-\chi_{d|a}^{{\rm net}}})
-\chi_{a|d}^{{\rm net}}\:\chi_{d|a}^{{\rm net}}\right.\nonumber \\
-\left.\left[(1-e^{-\chi_{d|a}^{{\rm net}}})\;
e^{-\chi_{a|d}^{{\rm net}}}-\chi_{d|a}^{{\rm net}}\right]
 (\chi_{a|d}^{{\rm net}}-
 \lambda_{a(j)}\,\chi_{a\mathbb{P}}^{\mathbb{P}}(Y-y_{1},|\vec{b}-\vec{b}_{1}|)\right\}
 =2\chi_{ad(jl)}^{{\rm enh}}(Y,b)\,.\label{chi-tree-tot}\end{eqnarray}
 
\begin{figure*}[htb]
\begin{center}\includegraphics[%
  width=15.cm,
  height=6cm]{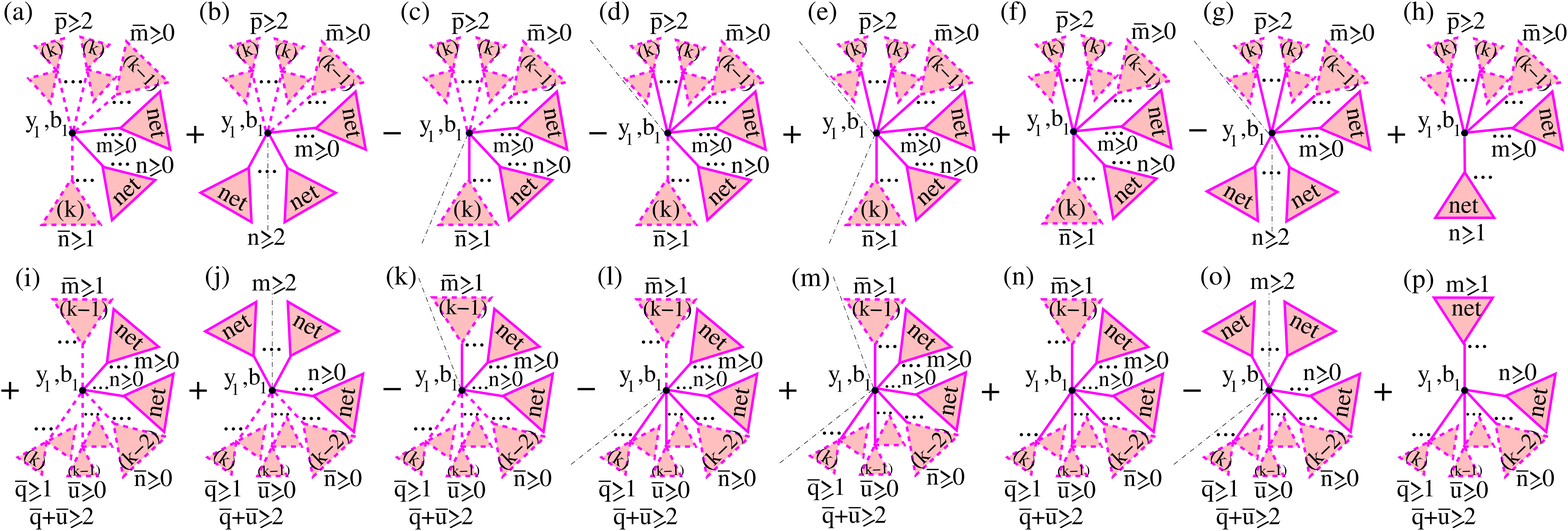}\end{center}
\vspace{-3mm}
\caption{Cut enhanced diagrams with non-tree-like topology of cut Pomerons.
\label{zz-cuts-fig}}
\end{figure*}

However, the unitarity requires  the sum of {\it all} the cuts of the diagrams
of Fig.~\ref{enh-full} to be equal to twice the imaginary part of the
 elastic scattering contribution, i.e.~to $2\chi_{ad(jl)}^{{\rm enh}}$.
Thus, the contributions of all cuts of non-tree (zigzag)
type should precisely cancel each other. To verify that, we can construct
the complete set of corresponding cut diagrams replacing
in Figs.~\ref{treecut1-fig} and \ref{treecut2-fig} some contributions
$\bar{\chi}_{a(j)|d(l)}^{{\rm fan}}$ and $\tilde{\chi}_{a(j)|d(l)}^{{\rm fan}}$
($\bar{\chi}_{d(l)|a(j)}^{{\rm fan}}$ and $\tilde{\chi}_{d(l)|a(j)}^{{\rm fan}}$
) by $\bar{\chi}_{a(j)|d(l)}^{{\rm zz(k)}}$ and $\tilde{\chi}_{a(j)|d(l)}^{{\rm zz(k)}}$
($\bar{\chi}_{d(l)|a(j)}^{{\rm zz(k)}}$ and $\tilde{\chi}_{d(l)|a(j)}^{{\rm zz(k)}}$),
whereas the others -- by $\bar{\chi}_{a(j)|d(l)}^{{\rm net(k-1)}}$ and 
$\tilde{\chi}_{a(j)|d(l)}^{{\rm net(k-1)}}$
($\bar{\chi}_{d(l)|a(j)}^{{\rm net(k-1)}}$ and $\tilde{\chi}_{d(l)|a(j)}^{{\rm net(k-1)}}$),
starting from $k=2$, etc. Using the representation of Fig.~\ref{zigcut2-fig}
for $2\bar{\chi}_{a(j)|d(l)}^{{\rm zz}(k)}-2\tilde{\chi}_{a(j)|d(l)}^{{\rm zz}(k)}$
(similarly for $2\tilde{\chi}_{a(j)|d(l)}^{{\rm zz}(k)}$)
to correct for double counts in the same way as above for the tree-like cut
diagrams, we obtain the set of graphs of Fig.~\ref{zz-cuts-fig}.%
There, the diagrams (a)--(c) contain   $\bar p\geq 2$ $k$-th order 
projectile zigzag-like cut {}``net fans''; this gives a factor
$(e^{2\bar \chi_{a(j)|d(l)}^{{\rm zz}(k)}}-1-2\bar \chi_{a(j)|d(l)}^{{\rm zz}(k)})$,
which is equal  zero due to  (\ref{zz=0}). Similarly, the graphs (i)--(k)
have $\bar q\geq 1$ $k$-th order target zigzag-like cut {}``net fans'', 
which gives $(e^{2\bar \chi_{d(l)|a(j)}^{{\rm zz}(k)}}-1)=0$. 
The contributions of the graphs (e) and (f) are equal up to a sign and cancel
each other; the same applies to the diagrams (m) and (n). Finally, the graphs
(d), (g), (h) give together
 \begin{eqnarray}
G\int_{0}^{Y}\!\! dy_{1}\!\int\! d^{2}b_{1}\;
(e^{\tilde \chi_{a|d}^{{\rm zz}(k)}}-1-\tilde \chi_{a|d}^{{\rm zz}(k)})
 \;e^{\tilde\chi_{a|d}^{{\rm net}(k-1)}-\chi_{a|d}^{{\rm net}}}\nonumber \\
\times  \left[-(e^{2\bar\chi_{d|a}^{{\rm net}(k)}}-1)\;
 e^{-2\chi_{d|a}^{{\rm net}}}
-(1-e^{-\chi_{d|a}^{{\rm net}}})^2
+2 (1-e^{-\chi_{d|a}^{{\rm net}}}) \right]\!,
\label{cancel}
\end{eqnarray}
where the expression in the square brackets vanishes due to (\ref{equiv1}).
Similarly one demonstrates the cancellation for the graphs (l), (o), and (p).
 This completes the proof of the $s$-channel unitarity of
the approach.
 It is worth stressing  that we obtained a  cancellation
for the contributions of non-tree type cut diagrams of Fig.~\ref{zz-cuts-fig}
to the total cross section, not  to inclusive particle spectra;
 such graphs   have to be taken into consideration
in inelastic event generation procedures.

\begin{figure}[t]
\begin{center}\includegraphics[%
  width=8cm,
  height=4cm]{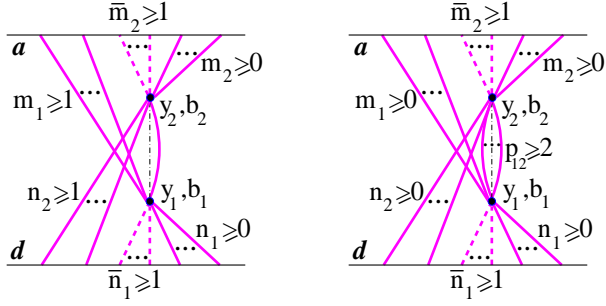}\end{center}
\vspace{-3mm}
\caption{Lowest order contributions to the double high mass 
diffraction cross section: net-like diagrams (left)
and loop graphs (right).
\label{gap-fig}}
\end{figure}

It is noteworthy that the $s$-channel unitarity is still violated 
in the described scheme in certain parts of the kinematic space,
which is the price for neglecting Pomeron loop contributions. 
For example, one  obtains here a negative contribution
for double high mass diffraction (central  rapidity gap)
 cross section $\sigma _{ad}^{\rm DD}$.
Indeed, dominant contribution to the process comes from hadron-hadron
scattering at relatively large impact parameters, where the RGS
probability is not too small, and, due to the smallness of the 
triple-Pomeron coupling $r_{3{\mathbb{P}}}$, is given by the graphs
 of Fig.~\ref{gap-fig}~(left),%
 with only two multi-Pomeron vertices.
 Thus, for $\sigma _{ad}^{\rm DD}$ one obtains 
\begin{eqnarray}
\sigma _{ad}^{\rm DD}(s,y_1,y_2)\simeq
-\frac{G}{2}\sum _{j,l}C_{a(j)}\, C_{d(l)}\int\! d^{2}b\,d^{2}b_{1}\,d^{2}b_{1}\;
 \chi_{\mathbb{PP}}^{\mathbb{P}}(y_{2}-y_{1},|\vec{b}_{2}-\vec{b}_{1}|)\nonumber \\
\times (1-e^{-2\lambda_{a(j)}\,\chi_{a\mathbb{P}}^{\mathbb{P}}(Y-y_{2},|\vec{b}-\vec{b}_{2}|)})\;
(1-e^{-\lambda_{d(l)}\,\chi_{d\mathbb{P}}^{\mathbb{P}}(y_{2},b_{2})})\;
e^{-\lambda_{d(l)}\,\chi_{d\mathbb{P}}^{\mathbb{P}}(y_{2},b_{2})}\;
(1-e^{-2\lambda_{d(l)}\,\chi_{d\mathbb{P}}^{\mathbb{P}}(y_{1},b_{1})})\nonumber \\
\times (1-e^{-\lambda_{a(j)}\,\chi_{a\mathbb{P}}^{\mathbb{P}}(Y-y_{1},|\vec{b}-\vec{b}_{1}|)})\;
e^{-\lambda_{a(j)}\,\chi_{a\mathbb{P}}^{\mathbb{P}}(Y-y_{1},|\vec{b}-\vec{b}_{1}|)}\;
S^{\rm RG}_{ad(jl)}(Y,b,y_1,y_2)\,,
\label{chi-LRG-1}
\end{eqnarray}
where $S^{\rm RG}_{ad(jl)}(Y,b,y_1,y_2)$ is the (positively defined) RGS factor,
 i.e.~the probability that additional re-scattering processes produce no secondary particles 
 in the rapidity interval $(y_1,y_2)$. 

\section{Pomeron loops}
\begin{figure}[t]
\begin{center}\includegraphics[%
  width=12cm,
  height=4cm]{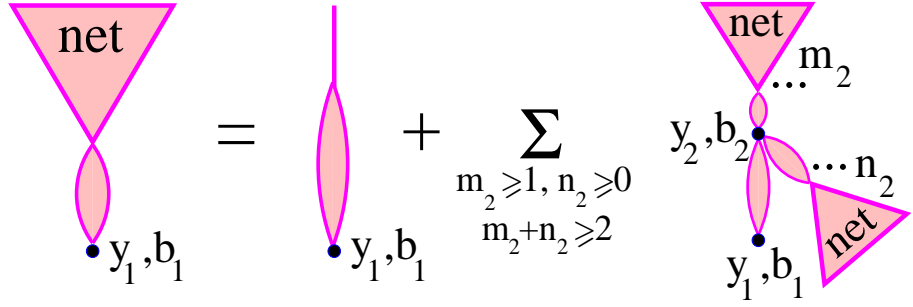}\end{center}
\vspace{-3mm}
\caption{Generalized {}``net fan'' contribution generates Pomeron nets,
whose  neighboring ``cells'' are connected 
by $t$-channel sequences of Pomerons and Pomeron loops.
\label{net-loop-fig}}
\end{figure}

\begin{figure}[t]
\begin{center}\includegraphics[%
  width=8cm,
  height=6cm]{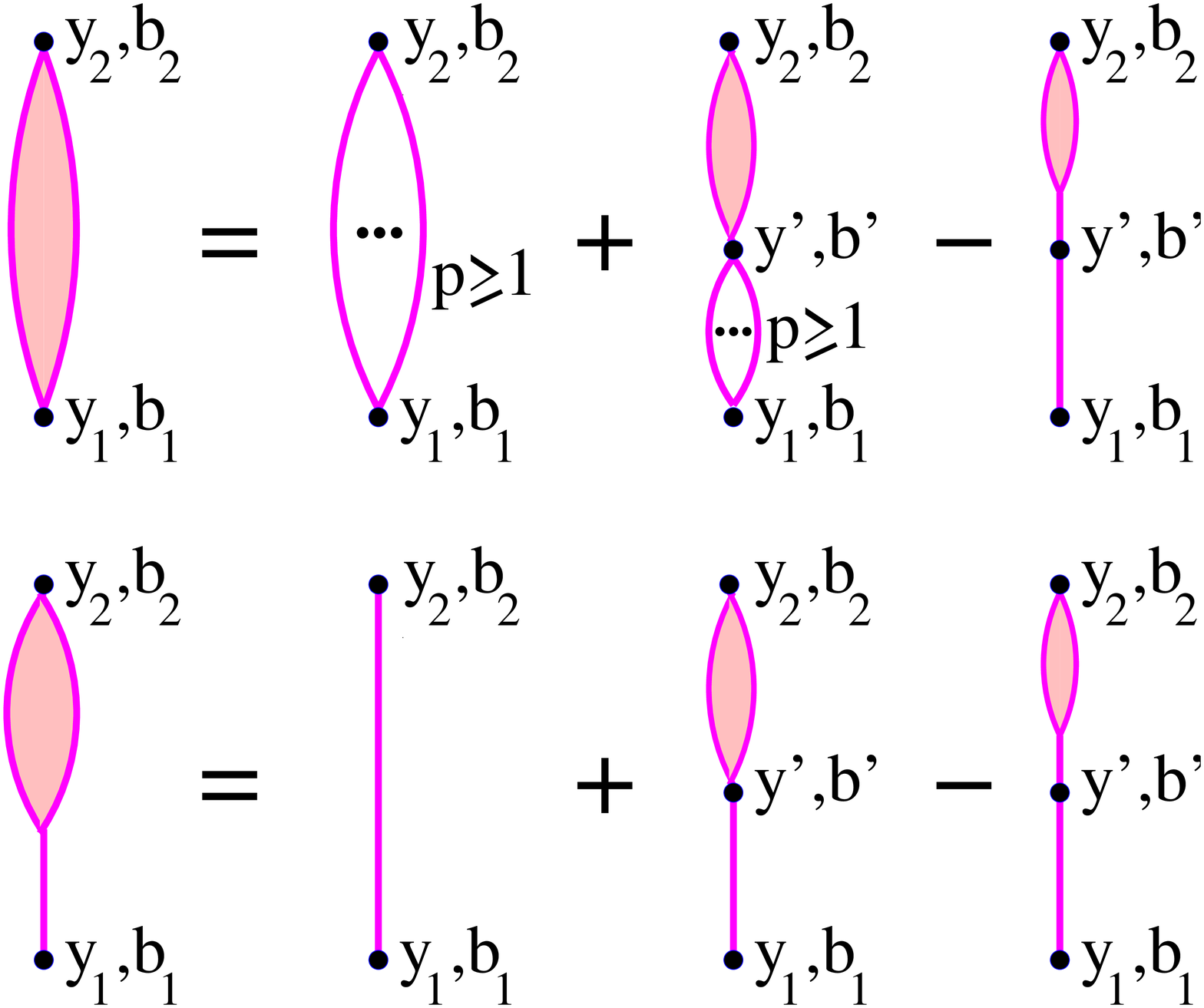}\end{center}
\vspace{-3mm}
\caption{Recursive representation for the contributions of Pomeron loop
sequences $\chi_{\mathbb{PP}}^{\rm loop}$, $\chi_{\mathbb{PP}}^{\rm loop(1)}$,
 exchanged between  the vertices  $(y_{1},b_{1})$ and  $(y_{2},b_{2})$.
\label{loop-fig}}
\end{figure}

The above-described procedure can be easily generalized to include simple loop
contributions, replacing single Pomerons connecting neighboring ``cells'' of Pomeron ``nets''
by $t$-channel sequences of Pomerons and Pomeron loops. To this end, one can modify the 
definition (\ref{net-fan}) of the {}``net fan'' contributions $\chi_{a(j)|d(l)}^{{\rm net}}$,
as shown in Fig.~\ref{net-loop-fig},%
 i.e.
 \begin{eqnarray}
\chi_{a(j)|d(l)}^{{\rm net}}(y_{1},\vec{b}_{1}|Y,\vec{b})
=\lambda_{a(j)}\,\chi_{a\mathbb{P}}^{\rm loop}(y_{1},b_{1})
+G\int_{0}^{y_{1}}\! dy_{2}\int\! d^{2}b_{2}\;
\chi_{\mathbb{PP}}^{\rm loop}(y_{1}-y_{2},|\vec{b}_{1}-\vec{b}_{2}|)\nonumber \\
\times\left\{ (1-e^{-\chi_{a(j)|d(l)}^{{\rm net}}\!(y_{2},\vec{b}_{2}|Y,\vec{b})})
\; e^{-\chi_{d(l)|a(j)}^{{\rm net}}\!(Y-y_{2},\vec{b}-\vec{b}_{2}|Y,\vec{b})}
-\chi_{a(j)|d(l)}^{{\rm net}}(y_{2},\vec{b}_{2}|Y,\vec{b})\right\} \!,\label{net-loop}
\end{eqnarray}
where the contributions $\chi_{a\mathbb{P}}^{\rm loop}$ and $\chi_{\mathbb{PP}}^{\rm loop}$
of Pomeron loop sequences, exchanged between hadron $a$ and the vertex $(y_{1},b_{1})$,
respectively, between the vertices $(y_{1},b_{1})$ and $(y_{2},b_{2})$, 
are defined via the recursive representations of 
Figs.~\ref{loop-leg-fig} and \ref{loop-fig}:%
\begin{figure}[t]
\begin{center}\includegraphics[%
  width=8cm,
  height=6cm]{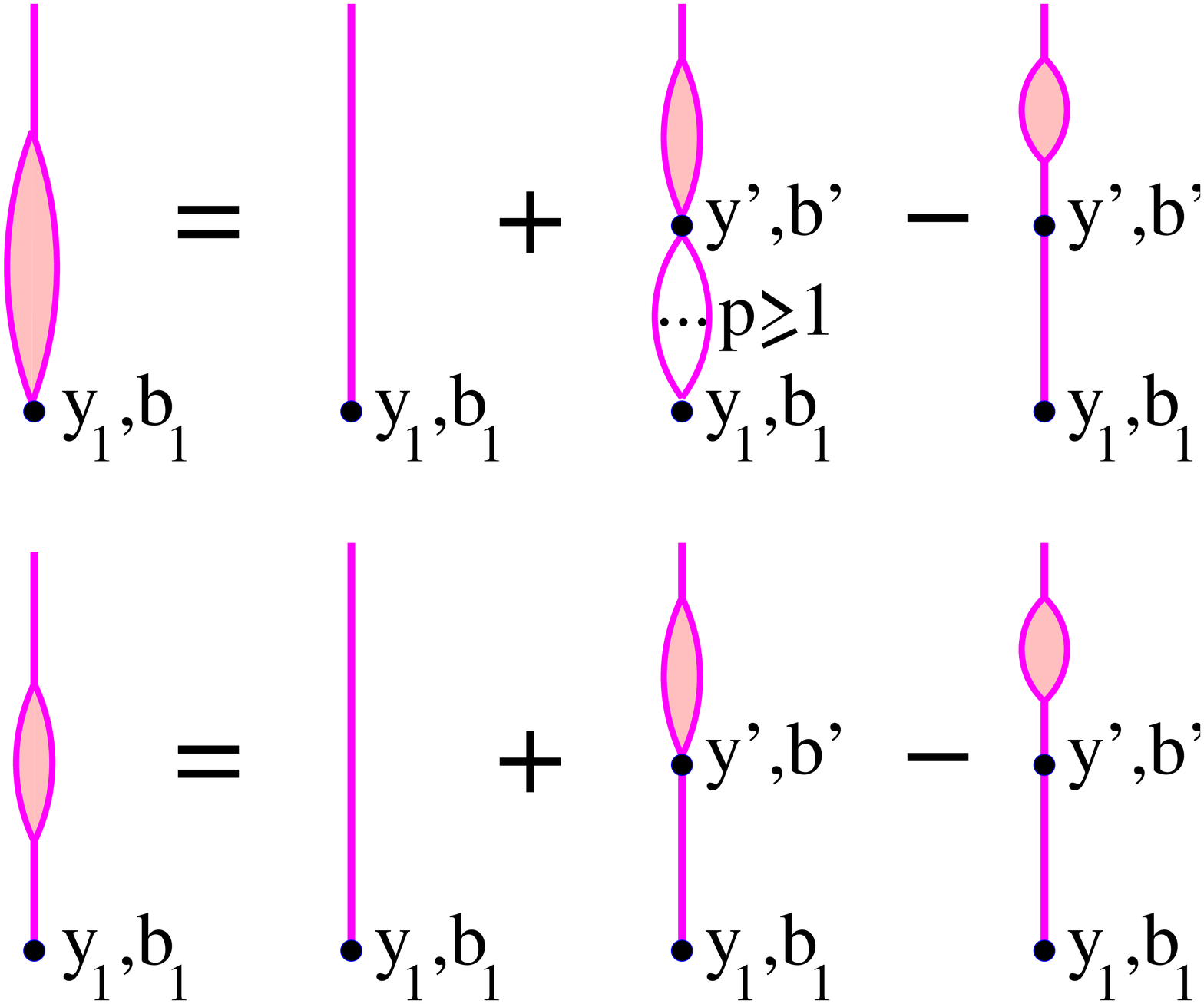}\end{center}
\vspace{-3mm}
\caption{Recursive representation for the contributions of Pomeron loop
sequences $\chi_{a\mathbb{P}}^{\rm loop}$, $\chi_{a\mathbb{P}}^{\rm loop(1)}$,
 exchanged between hadron $a$ and the vertex  $(y_{1},b_{1})$.
\label{loop-leg-fig}}
\end{figure}
 \begin{eqnarray}
\chi_{a\mathbb{P}}^{\rm loop}(y_{1},b_{1})=\chi_{a\mathbb{P}}^{\mathbb{P}}(y_{1},b_{1})\nonumber \\
+G\int_{0}^{y_{1}}\! dy'\int\! d^{2}b'\,
\left[ (1-e^{-\chi_{\mathbb{PP}}^{\mathbb{P}}(y_{1}-y',|\vec{b}_{1}-\vec{b}'|)})\;
\chi_{a\mathbb{P}}^{\rm loop}(y',b')
-\chi_{\mathbb{PP}}^{\mathbb{P}}(y_{1}-y',|\vec{b}_{1}-\vec{b}'|)\;
\chi_{a\mathbb{P}}^{\rm loop(1)}(y',b')\right]\label{chi-loop-leg}\\
\chi_{a\mathbb{P}}^{\rm loop(1)}(y_{1},b_{1})=\chi_{a\mathbb{P}}^{\mathbb{P}}(y_{1},b_{1})
+G\int_{0}^{y_{1}}\! dy'\int\! d^{2}b'\;
 \chi_{\mathbb{PP}}^{\mathbb{P}}(y_{1}-y',|\vec{b}_{1}-\vec{b}'|)
\left[\chi_{a\mathbb{P}}^{\rm loop}(y',b')-
\chi_{a\mathbb{P}}^{\rm loop(1)}(y',b')\right]\label{chi-loop-leg(1)}\\
\chi_{\mathbb{PP}}^{\rm loop}(y_{2}-y_{1},|\vec{b}_{2}-\vec{b}_{1}|)=
1-e^{-\chi_{\mathbb{PP}}^{\mathbb{P}}(y_{2}-y_{1},|\vec{b}_{2}-\vec{b}_{1}|)}
+G\int_{y_1}^{y_{2}}\! dy'\int\! d^{2}b'\,
\left[ (1-e^{-\chi_{\mathbb{PP}}^{\mathbb{P}}(y'-y_1,|\vec{b}'-\vec{b}_1|)})
\right.\nonumber \\
\times \left.\chi_{\mathbb{PP}}^{\rm loop}(y_2-y',|\vec{b}_2-\vec{b}'|)-
\chi_{\mathbb{PP}}^{\mathbb{P}}(y'-y_1,|\vec{b}'-\vec{b}_1|)\;
\chi_{\mathbb{PP}}^{\rm loop(1)}(y_2-y',|\vec{b}_2-\vec{b}'|)\right]\label{chi-loop}\\
\chi_{\mathbb{PP}}^{\rm loop(1)}(y_{2}-y_{1},|\vec{b}_{2}-\vec{b}_{1}|)=
\chi_{\mathbb{PP}}^{\mathbb{P}}(y_{2}-y_{1},|\vec{b}_{2}-\vec{b}_{1}|)\nonumber \\
+G\int_{y_1}^{y_{2}}\! dy'\int\! d^{2}b'\; 
\chi_{\mathbb{PP}}^{\mathbb{P}}(y'-y_1,|\vec{b}'-\vec{b}_1|)\;
\left[\chi_{\mathbb{PP}}^{\rm loop}(y_2-y',|\vec{b}_2-\vec{b}'|)-
\chi_{\mathbb{PP}}^{\rm loop(1)}(y_2-y',|\vec{b}_2-\vec{b}'|)\right]\!.
\end{eqnarray}
Here $\chi_{a\mathbb{P}}^{\rm loop(1)}$ and $\chi_{\mathbb{PP}}^{\rm loop(1)}$
are the contributions of such Pomeron loop sequences (exchanged between hadron $a$ 
and the vertex $(y_{1},b_{1})$, respectively, between the vertices $(y_{1},b_{1})$ 
and $(y_{2},b_{2})$), which start from a single Pomeron connected to the vertex
 $(y_{1},b_{1})$, as shown in   Figs.~\ref{loop-leg-fig}, \ref{loop-fig}.

\begin{figure}[htb]
\begin{center}\includegraphics[%
  width=14.5cm,
  height=4cm]{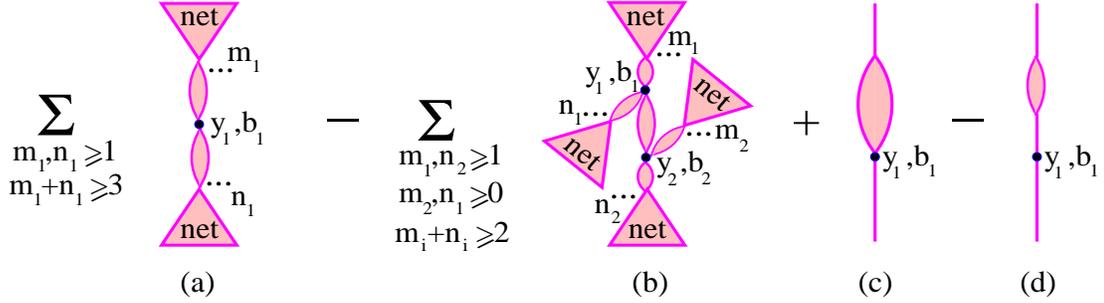}\end{center}
\vspace{-3mm}
\caption{Irreducible contributions of arbitrary Pomeron nets to elastic scattering
amplitude; neighboring net cells are connected by $t$-channel sequences of Pomerons
and Pomeron loops.
\label{enh-loop-fig}}
\end{figure}

Redefining in a similar way the fan diagram equation  (\ref{fan}),
one can literally repeat the analysis of Ref.~\cite{ost06} and obtain
the contribution of arbitrary Pomeron nets, with neighboring ``cells'' 
being connected  by $t$-channel loop sequences, in the form of 
Eq.~(\ref{chi-enh}), with the eikonal 
$\chi_{\mathbb{PP}}^{\mathbb{P}}(y_{1}-y_{2},|\vec{b}_{1}-\vec{b}_{2}|)$
being replaced by the corresponding  loop sequence contribution
$\chi_{\mathbb{PP}}^{\rm loop}(y_{1}-y_{2},|\vec{b}_{1}-\vec{b}_{2}|)$, 
as depicted in Fig.~\ref{enh-loop-fig}~(a),~(b).%

In addition, one has to consider an  exchange of a single $t$-channel
loop sequence between hadrons $a$ and $d$, as shown   
in Fig.~\ref{enh-loop-fig}~(c),~(d), such that each of the two hadrons is
coupled to a single Pomeron only. The complete eikonal contribution for the
considered class of enhanced diagrams is therefore
\begin{eqnarray}
\chi_{ad(jl)}^{{\rm enh}}(Y,b)=
G\int_{0}^{Y}\! dy_{1}\int\! d^{2}b_{1}\left\{ (1-e^{-\chi_{a|d}^{{\rm net}}(1)})
\,(1-e^{-\chi_{d|a}^{{\rm net}}(1)})
-\chi_{a|d}^{{\rm net}}(1)\;\chi_{d|a}^{{\rm net}}(1)\right.\nonumber \\
+\lambda_{a(j)}\,\lambda_{d(l)}\,\chi_{d\mathbb{P}}^{\mathbb{P}}(y_{1},b_{1})
\left[\chi_{a\mathbb{P}}^{\rm loop}(Y-y_{1},|\vec{b}-\vec{b}_{1}|)
-\chi_{a\mathbb{P}}^{\rm loop(1)}(Y-y_{1},|\vec{b}-\vec{b}_{1}|)\right]\nonumber \\
-G\int_{0}^{y_{1}}\! dy_{2}\int\! d^{2}b_{2}\;
\chi_{\mathbb{PP}}^{\rm loop}(y_{1}-y_{2},|\vec{b}_{1}-\vec{b}_{2}|)\;
 \left[(1-e^{-\chi_{a|d}^{{\rm net}}(1)})
\,e^{-\chi_{d|a}^{{\rm net}}(1)}-\chi_{a|d}^{{\rm net}}(1)\right]
\nonumber \\
\left.\times\left[(1-e^{-\chi_{d|a}^{{\rm net}}(2)})\,
e^{-\chi_{a|d}^{{\rm net}}(2)}-\chi_{d|a}^{{\rm net}}(2)\right]\right\}\!,
 \label{chi-enh-loop}
\end{eqnarray}
where we use the same abbreviations as in (\ref{chi-enh}).

\begin{figure*}[t]
\begin{center}\includegraphics[%
  width=15cm,
  height=7cm]{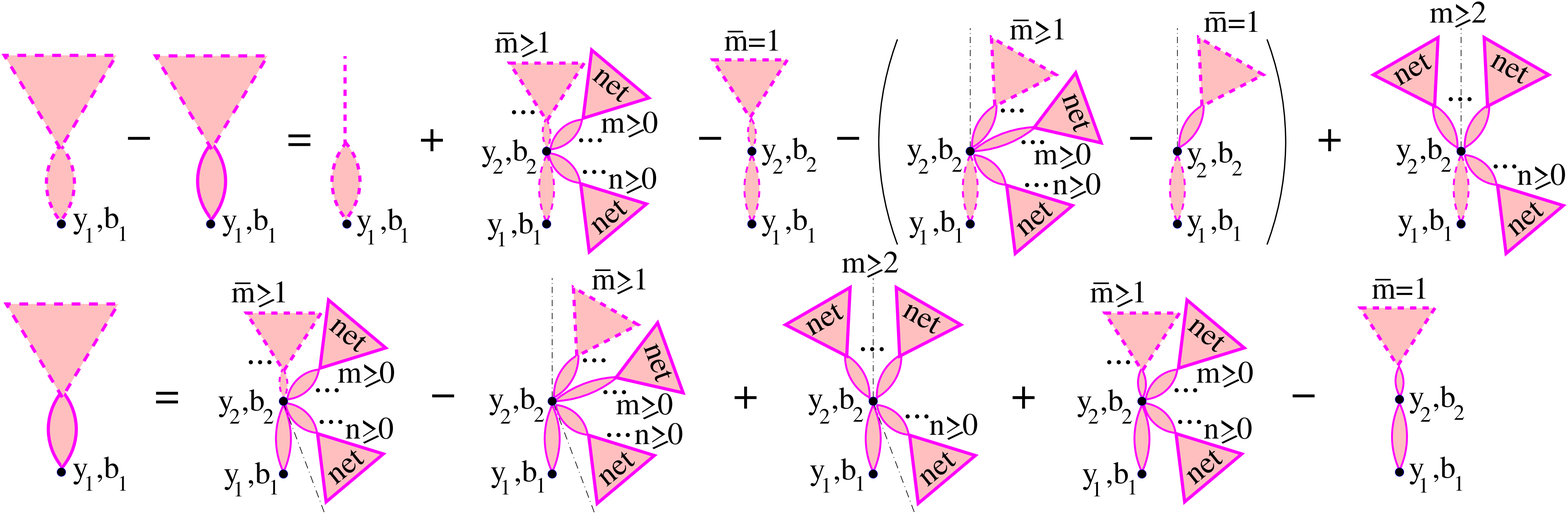}\end{center}
\vspace{-3mm}
\caption{Recursive equations for the contributions 
 $2\bar{\chi}_{a(j)|d(l)}^{{\rm fan}}-2\tilde{\chi}_{a(j)|d(l)}^{{\rm fan}}$, 
$2\tilde{\chi}_{a(j)|d(l)}^{{\rm fan}}$ of fan-like cuts  
 of generalized  {}``net fan'' graphs of Fig.~\ref{net-loop-fig}.
 \label{fan-loopc-fig}}
\end{figure*}

\begin{figure}[t]
\begin{center}\includegraphics[%
  width=9cm,
  height=6cm]{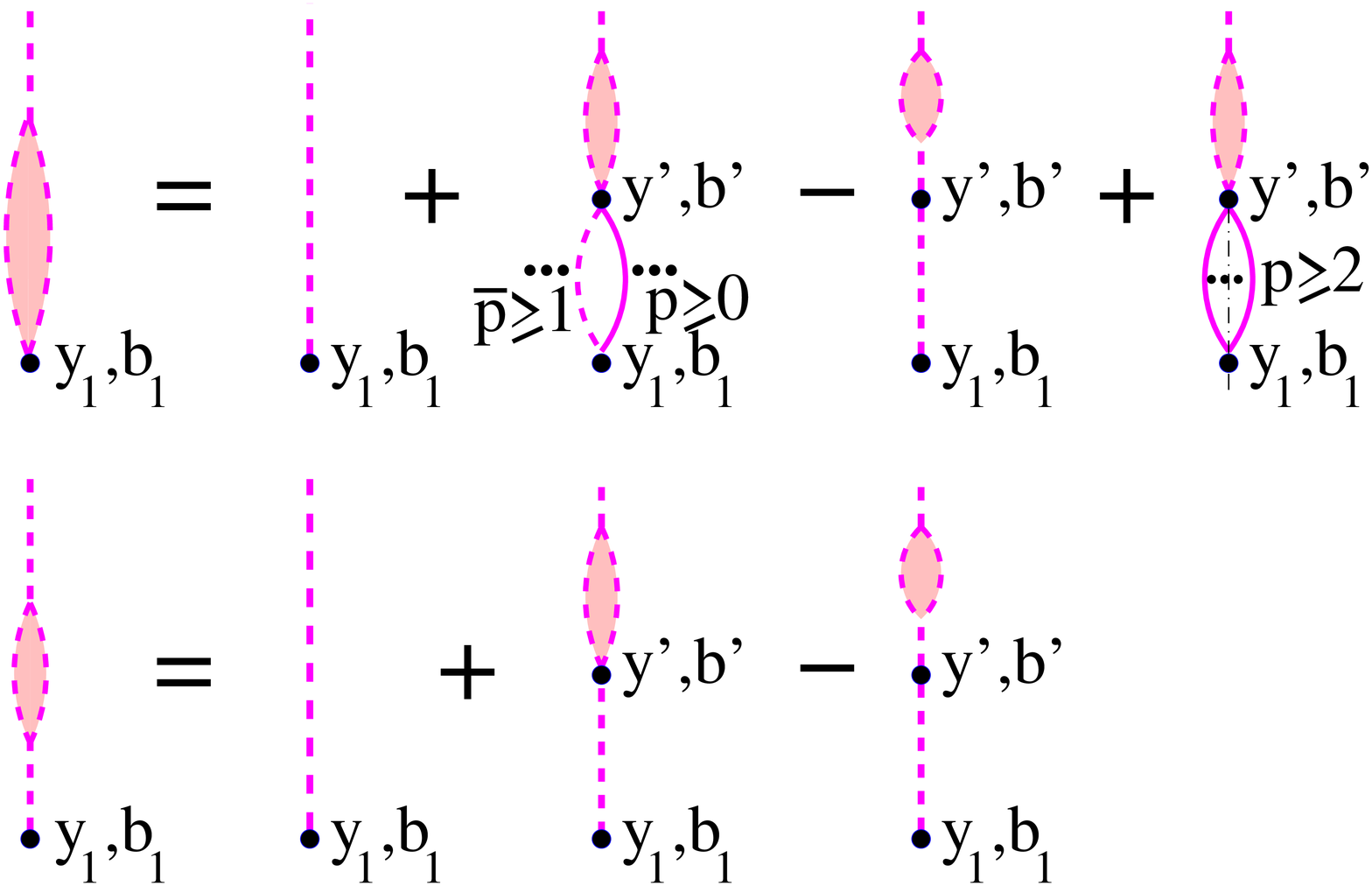}\end{center}
\vspace{-3mm}
\caption{Recursive representation for the AGK cuts of Pomeron loop
sequences, $\bar\chi_{a\mathbb{P}}^{\rm loop}$, $\bar\chi_{a\mathbb{P}}^{\rm loop(1)}$,
 exchanged between hadron $a$ and the vertex  $(y_{1},b_{1})$.
\label{loop-legc-fig}}
\end{figure}

\begin{figure}[t]
\begin{center}\includegraphics[%
  width=9cm,
  height=6cm]{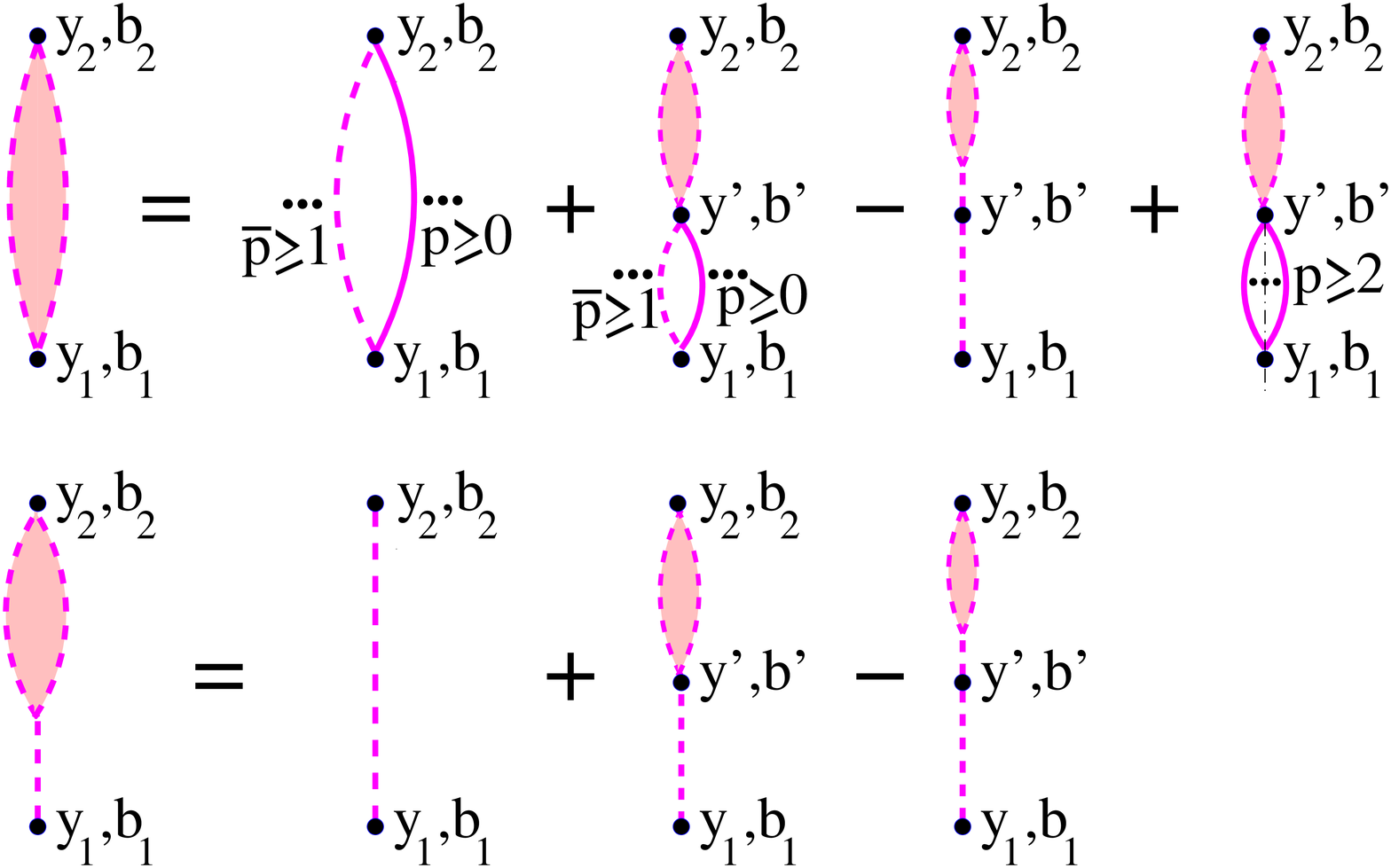}\end{center}
\vspace{-3mm}
\caption{Recursive representation for the AGK cuts of Pomeron loop
sequences, $\bar\chi_{\mathbb{PP}}^{\rm loop}$, $\bar\chi_{\mathbb{PP}}^{\rm loop(1)}$,
 exchanged between  the vertices  $(y_{1},b_{1})$ and  $(y_{2},b_{2})$.
\label{loopc-fig}}
\end{figure}

The analysis of unitarity cuts of the generalized scheme proceeds similarly to the
one described in Sections \ref{fans.sec} and \ref{enh.sec}. For the contribution of
fan-like cuts $2\bar{\chi}_{a(j)|d(l)}^{{\rm fan}}-2\tilde{\chi}_{a(j)|d(l)}^{{\rm fan}}$, 
$2\tilde{\chi}_{a(j)|d(l)}^{{\rm fan}}$ of  {}``net fan'' graphs of 
Fig.~\ref{net-loop-fig} one obtains the 
representations of Fig.~\ref{fan-loopc-fig}%
(c.f.~Figs.~\ref{fan-cut-fig},~\ref{fan-hole-fig}), i.e.
\begin{eqnarray}
2\bar{\chi}_{a(j)|d(l)}^{{\rm fan}}(y_{1},\vec{b}_{1}|Y,\vec{b})
-2\tilde{\chi}_{a(j)|d(l)}^{{\rm fan}}(y_{1},\vec{b}_{1}|Y,\vec{b})
=2\lambda_{a(j)}\,\bar{\chi}_{a\mathbb{P}}^{\rm loop}(y_{1},b_{1})\nonumber \\
+G\int_{0}^{y_{1}}\! dy_{2}\int\! d^{2}b_{2}\;
\bar{\chi}_{\mathbb{PP}}^{\rm loop}
\left\{ \left[(e^{2\bar{\chi}_{a|d}^{{\rm fan}}}-1)\:
e^{-2\chi_{a|d}^{{\rm net}}-2\chi_{d|a}^{{\rm net}}}
-2\bar{\chi}_{a|d}^{{\rm fan}}\right]\right.\nonumber \\
\left.-2\left[(e^{\tilde{\chi}_{a|d}^{{\rm fan}}}-1)\:
e^{-\chi_{a|d}^{{\rm net}}-2\chi_{d|a}^{{\rm net}}}
-\tilde{\chi}_{a|d}^{{\rm fan}}\right]
+(1-e^{-\chi_{a|d}^{{\rm net}}})^{2}\:
e^{-2\chi_{d|a}^{{\rm net}}}\right\} \label{chi-cut-fanloop}\\
2\tilde{\chi}_{a(j)|d(l)}^{{\rm fan}}(y_{1},\vec{b}_{1}|Y,\vec{b})
=G\int_{0}^{y_{1}}\! dy_{2}\int\! d^{2}b_{2}\;
\chi_{\mathbb{PP}}^{\rm loop}
\left\{ (1-e^{-\chi_{d|a}^{{\rm net}}})\:
e^{-\chi_{d|a}^{{\rm net}}}
\left[(e^{2\bar{\chi}_{a|d}^{{\rm fan}}}-1)\:
e^{-2\chi_{a|d}^{{\rm net}}}\right.\right.\nonumber \\
\left.\left.-2\,(e^{\tilde{\chi}_{a|d}^{{\rm fan}}}-1)\:
e^{-\chi_{a|d}^{{\rm net}}}+(1-e^{-\chi_{a|d}^{{\rm net}}})^{2}\right]
+2\left[(e^{\tilde{\chi}_{a|d}^{{\rm fan}}}-1)\:
e^{-\chi_{a|d}^{{\rm net}}-\chi_{d|a}^{{\rm net}}}
-\tilde{\chi}_{a|d}^{{\rm fan}}\right]\right\} \!,\label{chi-cut-handleloop}
\end{eqnarray}
where the contributions of cut loop sequences $\bar{\chi}_{a\mathbb{P}}^{\rm loop}$,
$\bar{\chi}_{a\mathbb{P}}^{\rm loop(1)}$,
 $\bar{\chi}_{\mathbb{PP}}^{\rm loop}$, $\bar{\chi}_{\mathbb{PP}}^{\rm loop(1)}$
 satisfy the recursive equations of Figs.~\ref{loop-legc-fig} and \ref{loopc-fig}%
respectively. Clearly, one has
\begin{eqnarray}
\bar{\chi}_{a\mathbb{P}}^{\rm loop}(y_{1},b_{1})=
\chi_{a\mathbb{P}}^{\rm loop}(y_{1},b_{1}) \label{equiv-loop-legc}\\
\bar{\chi}_{a\mathbb{P}}^{\rm loop(1)}(y_{1},b_{1})=
\chi_{a\mathbb{P}}^{\rm loop(1)}(y_{1},b_{1}) \label{equiv-loop(1)-legc}\\
\bar{\chi}_{\mathbb{PP}}^{\rm loop}(y_{1}-y_{2},|\vec{b}_{1}-\vec{b}_{2}|)=
\chi_{\mathbb{PP}}^{\rm loop}(y_{1}-y_{2},|\vec{b}_{1}-\vec{b}_{2}|) \label{equiv-loopc} \\
\bar{\chi}_{\mathbb{PP}}^{\rm loop(1)}(y_{1}-y_{2},|\vec{b}_{1}-\vec{b}_{2}|)=
\chi_{\mathbb{PP}}^{\rm loop(1)}(y_{1}-y_{2},|\vec{b}_{1}-\vec{b}_{2}|)\,. \label{equiv-loopc(1)}
\end{eqnarray}

Thus, repeating literally the reasoning of  Sections \ref{fans.sec} and \ref{enh.sec},
for the complete set of tree-like AGK cuts of the graphs of Fig.~\ref{enh-loop-fig}~(a),~(b)
we can obtain the representation of Figs.~\ref{treecut1-fig}, \ref{treecut2-fig}, 
and \ref{treecut3-fig}, with the uncut and fan-like cut  {}``net fans'' 
being defined now as in Figs.~\ref{net-loop-fig} and \ref{fan-loopc-fig}
respectively, with the single cut Pomeron contribution $\chi_{d\mathbb{P}}^{\mathbb{P}}$
in Fig.~\ref{treecut3-fig}~(i),~(j) being replaced by the one of the cut loop sequence
$\bar{\chi}_{d\mathbb{P}}^{\rm loop}$, and with the t-channel sequences of cut and uncut
Pomerons $\chi_{a(j)|d(l)}^{\mathbb{P}_{{\rm cc}}}$ and $\chi_{a(j)|d(l)}^{\mathbb{P}_{{\rm uc}}}$ 
(depicted as broken Pomeron lines in Fig.~\ref{treecut3-fig}) being replaced
by the ones of loops  $\chi_{a(j)|d(l)}^{\rm{loop}_{{\rm cc}}}$,
 $\chi_{a(j)|d(l)}^{\rm{loop}_{{\rm uc}}}$. For the latter one obtains, similarly to
 (\ref{chi-a-cc}-\ref{chi-a-uc}),
\begin{eqnarray}
2\chi_{a(j)|d(l)}^{\rm{loop}_{{\rm cc}}}(y_{1},\vec b_{1}|Y,\vec{b})
=2\lambda_{a(j)}\,\chi_{a\mathbb{P}}^{\rm{loop}}(y_{1},b_{1})
+2G\int_{0}^{y_{1}}\! dy_2 \int\! d^{2}b_2 
\;\chi_{\mathbb{PP}}^{\rm{loop}}\nonumber \\
\times \left[
(e^{-2\chi_{a|d}^{{\rm net}}-2\chi_{d|a}^{{\rm net}}}-1)\,
(\chi_{a|d}^{\rm{loop}_{{\rm cc}}}+
\chi_{a|d}^{\rm{loop}_{{\rm uc}}})
- (e^{-\chi_{a|d}^{{\rm net}}-2\chi_{d|a}^{{\rm net}}}-1) \,
\chi_{a|d}^{\rm{loop}_{{\rm uc}}}\right]  \label{chi-legloop-cc}\\
2\chi_{a(j)|d(l)}^{\rm{loop}_{{\rm uc}}}(y_{1},\vec b_{1}|Y,\vec{b})
=2G\int_{0}^{y_{1}}\! dy_2\int\! d^{2}b_2
\;\chi_{\mathbb{PP}}^{\rm{loop}}\nonumber \\
\times\left[(1-e^{-\chi_{d|a}^{{\rm net}}})\,
e^{-2\chi_{a|d}^{{\rm net}}-\chi_{d|a}^{{\rm net}}}
 \,(\chi_{a|d}^{\rm{loop}_{{\rm cc}}}+
\chi_{a|d}^{\rm{loop}_{{\rm uc}}}) 
+(e^{-\chi_{a|d}^{{\rm net}}-2\chi_{d|a}^{{\rm net}}}-1)\,
\chi_{a|d}^{\rm{loop}_{{\rm uc}}}\right] \!.\label{chi-legloop-uc}
\end{eqnarray}

In a similar way one generalizes the definition of the  $k$-th order
 cut {}``net fan'' contributions $2\bar{\chi}_{a|d}^{{\rm net}(k)}$
 and obtains the representation of Fig.~\ref{zz-cuts-fig} for the
 complete set of zigzag-like cuts of the diagrams 
  of Fig.~\ref{enh-loop-fig}~(a),~(b). Finally, for the graphs
 of Fig.~\ref{enh-loop-fig}~(c),~(d)   the cutting procedure is
 trivial, yielding a convolution of the cut loop sequences 
 $\bar{\chi}_{a\mathbb{P}}^{\rm loop}$ and $\bar{\chi}_{a\mathbb{P}}^{\rm loop(1)}$
 of Fig.~\ref{loop-legc-fig} with the cut Pomeron eikonal $\chi_{d\mathbb{P}}^{\mathbb{P}}$:
 \begin{eqnarray}
2G\int_{0}^{Y}\! dy_{1}\int\! d^{2}b_{1}\;
\lambda_{a(j)}\,\lambda_{d(l)}\,\chi_{d\mathbb{P}}^{\mathbb{P}}(y_{1},b_{1})
\left[\bar{\chi}_{a\mathbb{P}}^{\rm loop}(Y-y_{1},|\vec{b}-\vec{b}_{1}|)
-\bar{\chi}_{a\mathbb{P}}^{\rm loop(1)}(Y-y_{1},|\vec{b}-\vec{b}_{1}|)\right]\!.
 \label{chi-enh-loop(1)}
\end{eqnarray}

 The described  generalization of the scheme appears to be
 sufficient to cure the above-mentioned problems with the 
 violation of the $s$-channel unitarity   in certain  kinematic regions.
 In the considered case of double high mass diffraction, in addition
 to the graph of  Fig.~\ref{gap-fig}~(left) one obtains now the loop
 diagram of  Fig.~\ref{gap-fig}~(right), such that the summary contribution
 becomes
\begin{eqnarray}
\sigma _{ad}^{\rm DD}(s,y_1,y_2)\simeq
\frac{G}{4}\sum _{j,l}C_{a(j)}\, C_{d(l)}\int\! d^{2}b\,d^{2}b_{1}\,d^{2}b_{1}\;
 (1-e^{-2\lambda_{a(j)}\,\chi_{a\mathbb{P}}^{\mathbb{P}}(Y-y_{2},|\vec{b}-\vec{b}_{2}|)})
 \nonumber \\
\times (1-e^{-2\lambda_{d(l)}\,\chi_{d\mathbb{P}}^{\mathbb{P}}(y_{1},b_{1})})\;
(1-e^{-\chi_{\mathbb{PP}}^{\mathbb{P}}(y_{2}-y_{1},|\vec{b}_{2}-\vec{b}_{1}|)})\;
e^{-\lambda_{a(j)}\,\chi_{a\mathbb{P}}^{\mathbb{P}}(Y-y_{1},|\vec{b}-\vec{b}_{1}|)}\;
e^{-\lambda_{d(l)}\,\chi_{d\mathbb{P}}^{\mathbb{P}}(y_{2},b_{2})}\nonumber \\
\times \left[
(1-e^{-\chi_{\mathbb{PP}}^{\mathbb{P}}(y_{2}-y_{1},|\vec{b}_{2}-\vec{b}_{1}|)})\;
e^{-\lambda_{a(j)}\,\chi_{a\mathbb{P}}^{\mathbb{P}}(Y-y_{1},|\vec{b}-\vec{b}_{1}|)}\;
e^{-\lambda_{d(l)}\,\chi_{d\mathbb{P}}^{\mathbb{P}}(y_{2},b_{2})}
\right. \nonumber \\
-\left. 2\,(1-e^{-\lambda_{a(j)}\,\chi_{a\mathbb{P}}^{\mathbb{P}}(Y-y_{1},|\vec{b}-\vec{b}_{1}|)})\;
(1-e^{-\lambda_{d(l)}\,\chi_{d\mathbb{P}}^{\mathbb{P}}(y_{2},b_{2})})\right]
S^{\rm RG}_{ad(jl)}(Y,b,y_1,y_2)\,.
\label{chi-LRG-2}
\end{eqnarray}
 In the region of large impact parameters, which gives the dominant contribution to
 (\ref{chi-LRG-2}), either $\chi_{a\mathbb{P}}^{\mathbb{P}}(Y-y_{1},|\vec{b}-\vec{b}_{1}|)$
 or/and $\chi_{d\mathbb{P}}^{\mathbb{P}}(y_{2},b_{2})$ is small. Thus, the expression in 
 the square brackets reduces to
\begin{eqnarray}
\chi_{\mathbb{PP}}^{\mathbb{P}}(y_{2}-y_{1},|\vec{b}_{2}-\vec{b}_{1}|)
-2\,\lambda_{a(j)}\,\lambda_{d(l)}\,
\chi_{a\mathbb{P}}^{\mathbb{P}}(Y-y_{1},|\vec{b}-\vec{b}_{1}|)\,
\chi_{d\mathbb{P}}^{\mathbb{P}}(y_{2},b_{2})>0\nonumber \\
\end{eqnarray}
and assures a positive result for $\sigma _{ad}^{\rm DD}$.
 A systematic analysis of hadronic final states, obtained in the described
 scheme, will be presented elsewhere \cite{ost08}.

\section{Conclusions}
We proposed here a method for a re-summation of the {\it full} set of 
AGK-based unitarity cuts of a very general class of net-like enhanced Pomeron diagrams.
This is the principal novelty of the present analysis compared to other related
works \cite{bond01,bor05,kmr07}, which have been restricted to investigations of 
contributions of particular, notably diffractive, final states only.
Though the main derivation has been performed for the class of non-loop net-like diagrams,
we have demonstrated that the method can be trivially generalized to include Pomeron loop
contributions. In the latter case, one simply considers neighboring cells of Pomeron
nets to be connected by (cut or uncut) $t$-channel sequences of Pomerons and Pomeron loops, 
rather than by single Pomeron exchanges; the same applies for the connections between
initial hadrons and the correspondingly neighboring net cells. In a similar way one can
include more general loop contributions \cite{ost08}.

Although the obtained expressions for the contributions of cut enhanced diagrams are
based on a particular eikonal ansatz (\ref{gmn}) for multi-Pomeron vertices, 
the corresponding diagrammatic representations, e.g.~of Figs.~\ref{enh-full}, 
\ref{freve}, \ref{fan-cut-fig}, \ref{fan-hole-fig},
\ref{treecut1-fig}, \ref{treecut2-fig}, \ref{treecut3-fig}, \ref{zz-cuts-fig},
are of more general character and remain applicable for arbitrary parameterizations
of multi-Pomeron vertices.

It is noteworthy that current analysis does not depend on a particular
parameterization for the Pomeron exchange amplitude and can be extended
 for a phenomenological description of {}``hard'' partonic processes \cite{ost06a}. 
 In principle, the proposed method can be also applied  in the perturbative
 BFKL Pomeron framework. However, one should keep in mind that the principal
 assumption of the present analysis was that  the AGK cutting
 rules remain valid, in particular, that multi-Pomeron vertices remain unmodified
 by the cutting procedure. The fact that the AGK rules are not proven in QCD,
 with some  deviations from the AGK prescriptions already reported in literature \cite{nik06},
 implies that the method may have to be significantly modified, when employed in 
 the BFKL Pomeron calculus. On the other hand, recent investigations indicate
 that the AGK picture still remains a reasonably good approximation in the pQCD
 framework \cite{bart06}.

 The obtained results open the way for a consistent implementation
of the RFT in hadronic MC models. Details of the corresponding
procedure will be discussed elsewhere \cite{ost08}.
On the other hand, the scheme can be applied for calculations of total
and diffractive hadronic  cross sections and of rapidity gap survival probabilities,
a preliminary analysis already  reported in \cite{ost06a}.

\end{document}